# 2D MXene integrated optical polarizers based on $TiO_2$ with high performance

*David J. Moss**

Optical Sciences Centre
Swinburne University of Technology
Melbourne, Victoria 3122, Australia
E-mail: *dmoss@swin.edu.au*

## Abstract

Polarization control is essential for modern optical systems. Recently, two-dimensional (2D) materials featuring highly anisotropic absorption across broad wavelength regions have emerged as a promising solution for realizing optical polarizers with operation bandwidths far exceeding those achievable using bulk materials. Despite this, existing 2D-material-based optical polarizers face critical limitations in simultaneously delivering high polarizer figures of merit (FOMs), robust power endurance, and scalable fabrication, hindering their practical deployment beyond laboratory. Here, we overcome this bottleneck by demonstrating optical polarizers through integrating 2D MXene films onto silicon photonic devices and subsequently converting them in-situ into 2D titanium dioxide ($TiO_2$). The 2D $TiO_2$ exhibits significantly improved anisotropic absorption and thermal / chemical stability compared with MXene, enabling the hybrid polarizers to achieve a record high polarizer FOM of ~17.5 (among 2D-material-coated devices) and excellent power tolerance (stable under > 2 W coupled average power), together with a broad operation bandwidth (> 140 nm) and a high fabrication yield (~98%). We also demonstrate microring resonator polarizers achieving high polarization extinction ratios up to ~12 dB. With outstanding overall performance and strong capability for large-scale manufacturing, these polarizers represent a solid step towards industry-ready implementations of 2D-material-based optical polarizers.

## 1. Introduction

Polarization control underpins the functionality and performance of modern optical systems.[1, 2] Optical polarizers, which enable the selection of specific polarization states, have been widely used for this purpose.[3] In recent years, driven by the rapid advances in optical science and technology, there is rapidly growing demand for optical polarizers enabling efficient polarization selection over broad wavelength ranges,[4, 5] exposing the limitations of conventional bulk-material-based polarizers. Bulk-material-based optical polarizers have been implemented based on a range of device platforms, such as birefringent crystals,[6, 7] refractive prisms,[8, 9] optical fibers,[4, 10] and photonic integrated devices.[1, 3, 11-13] However, their operation bandwidths are usually restricted by material dispersion or frequency-dependent birefringence,[5, 14] and remain below 100 nm in most reported cases.

Owing to their highly anisotropic light absorption across broad wavelength regions,[4, 5, 15] two-dimensional (2D) materials with atomic thicknesses have emerged as a game changer to fundamentally address the bandwidth limitations facing bulk-material-based optical polarizers. On-chip integration of 2D materials to implement hybrid polarizers can not only inherit the advantages of photonic integrated devices, such as compact footprint, low power consumption, and mass producibility, but also enable operation bandwidths far exceeding those achievable with bulk-material-based polarizers.[16, 17] In contrast to bulk integrated materials that commonly face lattice mismatch issues,[18, 19] 2D materials feature dangling-bond-free surfaces and can be inherently adhered via van der Waals interactions.[20, 21] This facilitates their stress-free and flexible on-chip integration, providing significant potential for large-scale industrial manufacturing of hybrid integrated devices.

To date, 2D materials such as graphene, graphene oxide (GO), and transition metal dichalcogenides (TMDCs) have been on-chip integrated to implement hybrid optical polarizers, achieving broad operation bandwidths and high polarization-dependent loss (PDL). However, these polarizers remain limited in one form or another. For instance, graphene exhibits stronger material anisotropy than GO and TMDCs, resulting in higher polarizer figure of merits (FOMs). However, their practical implementation is hindered by high-temperature synthesis, cumbersome on-chip transfer, and limited control accuracy.[5, 22] GO has facile solution-based synthesis and film coating methods that enable transfer-free on-chip integration with precise control, but suffers from relatively low polarizer FOMs and limited power endurability due to GO reduction at high powers.[23, 24] TMDCs provide better power endurability but also face limitations arising from relatively low polarizer FOMs and the same fabrication constraints as graphene.[25] None of the current 2D-material-based integrated optical polarizers can simultaneously deliver high polarizer FOMs, robust power endurance, and scalable fabrication with precise control, which greatly limits their wider deployments and applications beyond laboratory.

In this work, we propose and experimentally demonstrate a new strategy to address existing bottlenecks in developing 2D-material-based integrated optical polarizers, enabling outstanding overall device performance and high readiness for industrial implementation. We first integrate 2D MXene films as precursors onto integrated silicon photonic devices using a transfer-free, layer-by-layer coating method with precise control, followed by their in-situ conversion into 2D titanium dioxide ($TiO_2$) films through a tailored two-step oxidation process. As an ultrathin metal oxide derived from MXene, 2D $TiO_2$ combines enhanced anisotropic light absorption, improved thermal / chemical stability, and scalable fabrication with precise control. Experimental results show that the integrated waveguide polarizers with 2D $TiO_2$ films simultaneously deliver a record high polarizer FOM of ~17.5 (among 2D-material-coated devices) and excellent power tolerance (stable under >2 W coupled average power), together with a broad operation bandwidth (> 140 nm) and a high fabrication yield (~98%). We also demonstrate $TiO_2$-coated microring resonator (MRR) polarizers achieving high polarization extinction ratios (*PER*s) up to ~12 dB. With combined performance metrics and strong capability for large-scale manufacturing, these 2D-material-based integrated optical polarizers establish a viable route towards industry-ready implementations and commercial deployment.

## 2. Selection of 2D material

In this work, we select 2D $Ti_3C_2T_x$, an important member of the MXene family that has been widely studied,[26] as a precursor for implementing high-performance optical polarizers. This choice is motivated by several considerations from both material and fabrication perspectives. First, MXene exhibits highly anisotropic light absorption over a very broad spectral bandwidth spanning from the visible to infrared regions,[27] and by engineering the material transformation to obtain $TiO_2$ as a derivative, the anisotropy can be further improved without compromising the response bandwidth. Second, we develop facile fabrication techniques that enable layer-by-layer, transfer-free MXene film coating with high scalability and precise control (as detailed in **Section 3**), which exhibit excellent compatibility with state-of-the-art integrated device platforms. Finally, the fabrication of 2D $TiO_2$ from MXene not only inherits the advantages for large-scale fabrication with precise control, but also leverages the intrinsically separated Ti layers in the MXene precursor to preserve structural anisotropy and improve polarization selectivity. The derived $TiO_2$ also exhibits very high thermal and chemical stability, enabling high-power tolerance and environmental stability for the hybrid polarizers (as demonstrated experimentally in **Section 5**).

**Figure 1a** illustrates the atomic structure of $Ti_3C_2T_x$, which consists of alternating stacks of Ti and C atomic layers.[26] Thick $Ti_3C_2T_x$ films are composed of multiple $Ti_3C_2$ layers, with the surfaces terminated by randomly distributed functional groups such as –O, –OH, and –F.

To derive $TiO_2$ from $Ti_3C_2T_x$, we develop a tailored two-step oxidation strategy for controlled material transformation, with thermally oxidized $Ti_3C_2T_x$ serving as an intermediate product. The process flow is illustrated in **Figure 1b**. For simplicity, $Ti_3C_2T_x$ and its thermally oxidized derivative are hereafter referred to as MXene and MXene oxide (MXO) in our following discussion, respectively. In the first step, MXene is oxidized to form MXO through thermal treatments at ~200 °C under ambient conditions. This can be readily achieved using hotplate heating and falls within the low-temperature regime of CMOS-compatible processing.[28] The second step involves further oxidation of MXO, accompanied by decarbonization and dehydration, to obtain $TiO_2$ as the final derivative, which is achieved through treatment with a piranha solution (composed of concentrated sulfuric acid ($H_2SO_4$, 98 wt%) and hydrogen peroxide ($H_2O_2$, 30 wt%) in a 3:1 volume ratio) – a cleaning method widely used in integrated device fabrication.[29] Further details and discussion of the two-step oxidation method for fabricating 2D $TiO_2$ films are provided in **Section 3**.

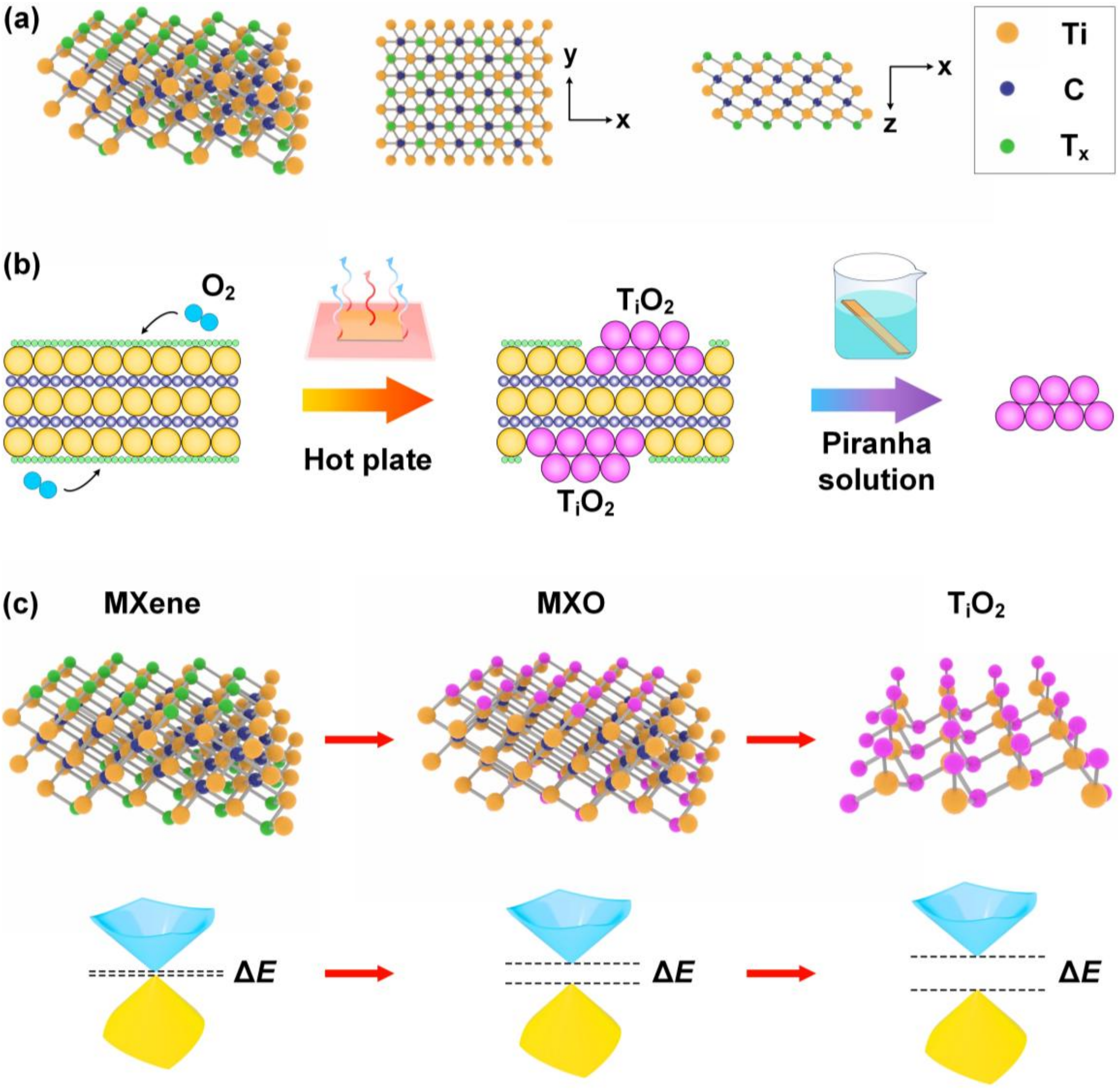

**Figure 1.** (**a**) Schematic atomic structure of MXene consisting of layered $Ti_3C_2$. (**b**) Schematic illustration of a two-step oxidation strategy for deriving titanium dioxide ($TiO_2$) from MXene in (**a**), with MXene oxide (MXO) serving as an intermediate product. (**c**) Schematics of atomic structures and optical bandgaps of MXene, MXO, and $TiO_2$ in (**b**).

**Figure 1c** illustrates the atomic structures of MXene, MXO, and $TiO_2$ in **Figure 1b**, together with Dirac cones illustrating their optical bandgaps. Compared with $Ti_3C_2$, which exhibits metallic behavior with a negligible bandgap,[26] MXene ($Ti_3C_2T_x$) shows a slightly opened bandgap induced by surface functionalization, typically on the order of ~0.1 eV.[26] The oxidation in the first step can disrupt the Ti–C network and alter the surface functionalization of $Ti_3C_2$ layers, thus leading to the formation of Ti–O bonds and an increased optical bandgap of ~2.3−2.9 eV for MXO.[30] After the second-step oxidation, $TiO_2$ exhibits a further increased optical bandgap of ~3.2 eV.[31] The large bandgap enables very low linear light absorption at visible and infrared wavelengths, with an in-plane material extinction coefficient ($k$) nearly two orders of magnitude lower than that of graphene.[31] Although $TiO_2$ exhibits relatively low light absorption, monolayer $TiO_2$ has a smaller thickness (~5 nm) than monolayer MXene (~6 nm) and MXO (~8 nm), and shows more significant anisotropy between in-plane and out-of-plane light absorption, which is beneficial for implementing 2D-material-based optical polarizers with high polarization selectivities.

Compared with 2D materials such as GO, $MoS_2$, and graphene, $TiO_2$ exhibits superior thermal and chemical stability owing to its metal oxide nature, which is highly desirable for realizing robust optical polarizers for practical applications. For thermal stability, $TiO_2$ can withstand temperatures of up to ~1000 °C.[32] In contrast, GO can be readily reduced at temperatures above 100 °C,[33] $MoS_2$ is prone to oxidation in ambient conditions at temperatures of ~300–400 °C,[34] and graphene is susceptible to oxidation in ambient conditions at temperatures of ~400–500 °C.[35] For chemical stability, graphene and GO cannot withstand either oxygen plasma or piranha solution treatment, $MoS_2$ can survive the former but not the latter. In contrast, $TiO_2$ remains stable after both treatments.

## 3. Device design and fabrication

**Figure 2a** shows the schematic of our proposed integrated waveguide polarizer consisting of a silicon photonic waveguide coated with a 2D $TiO_2$ film. The selection of the silicon photonic platform, which dominates integrated optics,[36, 37] is also based on the consideration of improving the industry readiness. The cross-sectional schematic of the hybrid waveguide is provided in **Figure 2b**. The 2D $TiO_2$ film is coated on the top surface of the silicon waveguide, enabling the interaction between the $TiO_2$ film and the waveguide evanescent field.

**Figure 2c** shows the corresponding transverse electric (TE) and transverse magnetic (TM) mode profiles (at 1550 nm) for the hybrid waveguide, which were simulated using commercial mode-solving software (COMSOL Multiphysics). In our simulation, the silicon waveguide had a top width of 400 nm and a thickness of 220 nm, and the thickness of the 2D $TiO_2$ film was 5 nm (corresponding to monolayer $TiO_2$). These parameters are consistent with our fabricated devices that exhibit optimized performance. The refractive index ($n$) and extinction coefficient ($k$) of $TiO_2$ for TE polarization were $n_{TE}$ = ~1.875 and $k_{TE}$ = ~0.0210, respectively. For TM polarization, the corresponding values were $n_{TM}$ = ~1.775 and $k_{TM}$ = ~0.0014. These values were obtained by fitting the measured transmission spectra of $TiO_2$-coated MRRs (see details in Note S1, Supporting Information). The simulated TE- and TM-polarized effective indices for the hybrid waveguide are ~2.171 + 2.165 × $10^{-4}$i and ~1.566 + 1.165 × $10^{-5}$i, respectively, which correspond to propagation loss (*PL*) of ~76.2 dB/cm and ~4.1 dB/cm. The large difference between them underpins the outstanding polarization selectivity of the hybrid waveguide polarizer.

From a physical perspective, the polarization selectivity originates from the anisotropic light absorption of the 2D $TiO_2$ film, which exhibits a large difference between in-plane and out-of-plane light absorption, as reflected by its $k_{TE}$ and $k_{TM}$. The silicon waveguide, although lacking anisotropic light absorption, provides a mature integrated device platform that enables efficient interaction between the evanescent optical field and the 2D $TiO_2$ film, where TE-polarized (in-plane) light experiences higher loss than the TM-polarized (out-of-plane) one, allowing the hybrid waveguide to operate as a TM-pass polarizer.

Polarization-dependent loss (*PDL*), which is defined as the difference in insertion loss (*IL*) between two orthogonal polarization states, has been widely used to quantitatively evaluate the polarization selectivity of an optical polarizer.[5, 16, 24] This is reasonable for polarizers based on bulk materials, as they typically exhibit very low loss for the high-transmission polarization state. However, for 2D-material-based waveguide polarizers, even the high-transmission polarization state can exhibit much higher *IL* than comparable uncoated waveguide. Although a high *PDL* can be readily achieved by increasing the coating length of 2D materials to accumulate a larger *IL* difference, this usually comes at the expense of significant excess insertion loss (*EIL*) induced by 2D materials. Here, the *EIL* is defined as the minimum *IL* introduced by the 2D material relative to the uncoated waveguide (which corresponds to the high-transmission polarization state). For instance, the GO and $MoS_2$ hybrid waveguide polarizers reported in Refs. [23] and [25] achieved high *PDL* of ~17 dB and ~21 dB, respectively, but at the cost of large *EIL* of ~10 dB and ~5 dB. Apparently, a high *EIL* results in extra power consumption and is undesirable for practical device applications. Therefore, the polarizer *FOM*,[5, 16] defined as the ratio of *PDL* to *EIL*, serves as a more comprehensive metric for evaluating the performance of 2D-material-based waveguide polarizers, particularly with regard to potential industry deployment.

In the following discussion, we will use both *PDL* and *FOM* to compare the performance of 2D-material-based waveguide polarizers. In addition to these two parameters, operation bandwidth and power tolerance are also key performance metrics for optical polarizers in practical applications. The former reflects the device's spectral versatility over a wide wavelength range and represents a key advantage of 2D-material-based optical polarizers over bulk-material-based counterparts. Whereas the latter reflects the device's robustness and reliability under high optical power, which is crucial for stable operation in practical and high-power systems.

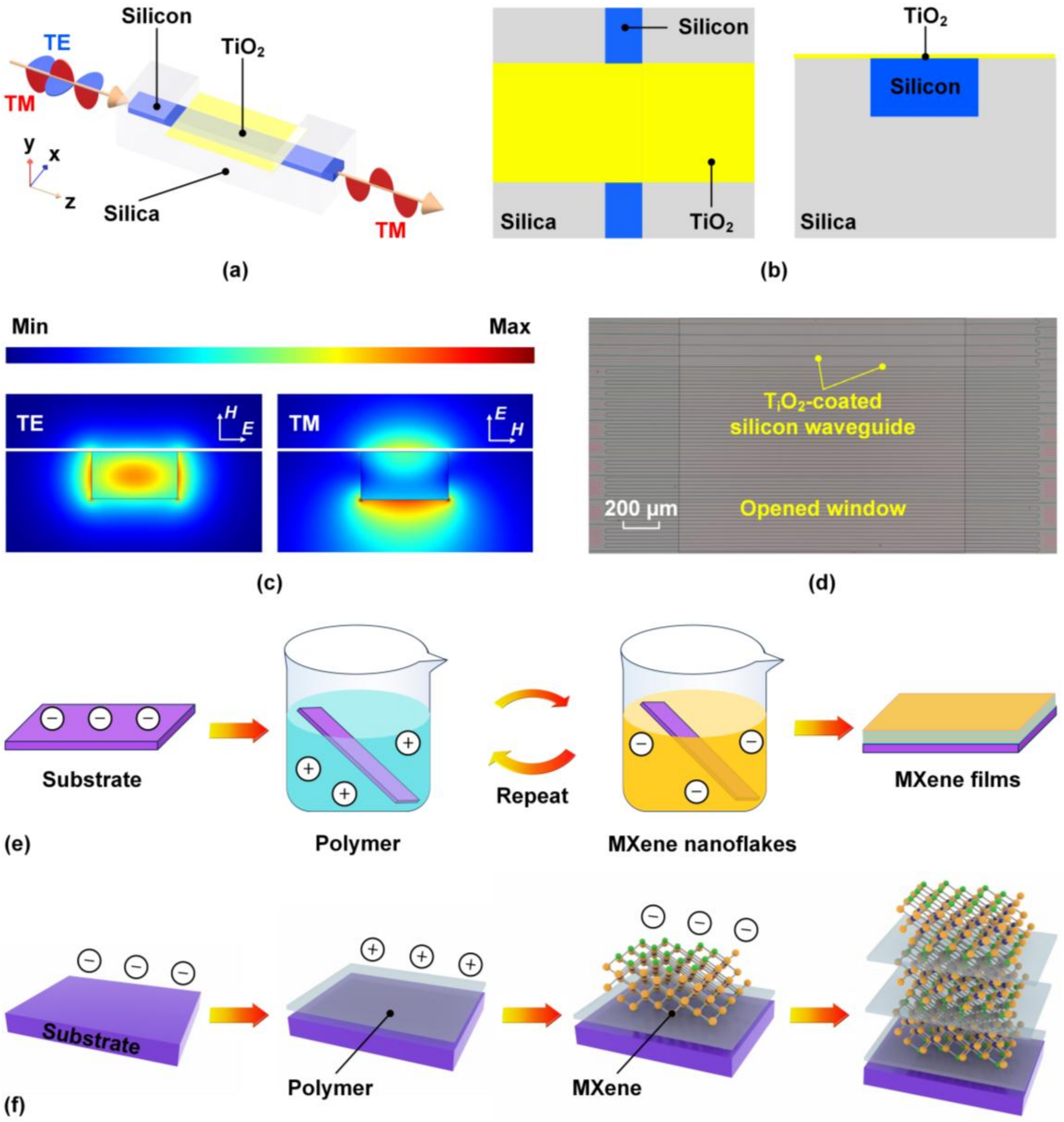

**Figure 2.** (**a**) Schematic illustration of an integrated waveguide polarizer based on a silicon photonic waveguide coated with a 2D $TiO_2$ film. (**b**) Schematic illustrations of top view and cross section of the hybrid waveguide. (**c**) Transverse electric (TE) and transverse magnetic (TM) mode profiles for the hybrid waveguide in (**b**). (**d**) Microscopic image of a silicon-on-insulator (SOI) chip coated with a 2D $TiO_2$ film. (**e**) Schematic illustration of layer-by-layer MXene film coating by using a solution-based self-assembly method. (**f**) Construction of a multi-layered MXene film during the self-assembly process.

**Figure 2d** shows a microscopic image of the fabricated hybrid waveguide polarizers on a silicon-on-insulator (SOI) chip. In our fabrication, we first fabricated the uncoated silicon waveguides on an SOI wafer using complementary metal-oxide-semiconductor (CMOS) compatible fabrication technologies. First, the silicon waveguides were formed by using 248-nm deep ultraviolet lithography followed by inductively coupled plasma etching. After fabricating the silicon waveguides, a silica upper cladding with a thickness of ~2 μm was deposited, and windows of different lengths were opened by etching the silica cladding down to the top surfaces of silicon waveguides for subsequent 2D film coating. The windows of varying lengths enable control of the interaction length between the 2D material and the waveguide evanescent field (*i.e.*, the 2D material coating length $L_c$ in our following discussion), which in turn determines the *PDL* of the hybrid polarizers.

The on-chip integration of 2D materials was performed after the fabrication of the silicon waveguides. We first prepared MXene nanoflakes by selectively etching the Al layers from the $Ti_3AlC_2$ MAX phase using an in situ LiF/HCl method,[38] followed by repeated washing and delamination to obtain a stable solution (see details in Note S2, Supporting Information). Next, 2D MXene films were integrated onto silicon waveguides through a solution-based self-assembly method that enabled transfer-free and layer-by-layer film coating with precise control of the film thickness. By repeating a four-step assembly process illustrated in **Figure 2e**, a multilayered film consisting of alternating MXene monolayers and polymer layers with oppositely charged surfaces was formed at room temperature via electrostatic forces, as shown in **Figure 2f**. Finally, the two-step oxidation process illustrated in **Figure 1b** was employed to in situ convert the 2D MXene precursor into 2D $TiO_2$. As shown in **Figure 2d**, the as-prepared $TiO_2$ film coated on an SOI chip exhibits high transmittance, smooth morphology, and high uniformity.

The two-step oxidation approach in **Figure 1b** allows a milder and better-controlled structural transformation compared with direct piranha-solution oxidation, thereby preserving material structural anisotropy and film uniformity while suppressing wrinkling and delamination. During the piranha treatment in the second step, the surfaces of MXO monolayers become highly hydroxylated (Ti–OH),[39] facilitating interlayer hydrogen bonding. Meanwhile, the decarbonization and dehydration of the polymer layers in this process promote Ti–OH condensation and Ti–O–Ti bond formation between adjacent layers, thus resulting in a layered $TiO_2$ structure. A similar process also occurs at the interface between the bottom $TiO_2$ layer and the silicon substrate, where the formation of robust interfacial Ti–O–Si bond[40, 41] enables firm adhesion of the $TiO_2$ films to silicon photonic devices. In our tests, the as-prepared $TiO_2$ films coated on silicon wafers remain intact after prolonged high-power sonication (~50 W/L for ~3 h).

2D films such as graphene and TMDCs are generally synthesized by chemical vapor deposition (CVD) processes at relatively high growth temperatures.[9, 11] Therefore, they are typically formed on metal substrates or foils, rather than on dielectric substrates for integrated platforms (*e.g.*, silicon and silica), thereby requiring subsequent transfer processes for on-chip integration.[42, 43] The transfer processes are usually cumbersome and face challenges in achieving high repeatability, uniform film coverage, accurate film placement, and effective removal of transfer-induced residues, which limit their wide deployment in large-scale manufacturing.[44-46] Although recent progress has been made in the direct growth of 2D materials on dielectric substrates,[47, 48] the inherently low surface energy of these substrates, along with the lack of catalyst-assisted growth, limits both the material quality and the capability for large-area growth.

In contrast, our coating method allows for transfer-free, layer-by-layer coating of 2D MXene films directly onto dielectric substrates, providing multiple compelling advantages for large-scale industrial manufacturing, including mild solution-based processing, high scalability, accurate control over the film thickness (via layer-by-layer deposition), high film uniformity (< 2 nm variation within a 4-inch wafer for monolayer film), and conformal coating onto dielectric substrates with structures of varying heights (enabled by the electrostatic forces). Building upon the self-assembly coating process, the in situ material transformation for fabricating 2D $TiO_2$ films not only preserves these advantages but also improves device performance with respect to the polarizer *FOM* and power tolerance (as experimentally demonstrated in **Section 5**), thereby further advancing its readiness for industrial deployment.

We note that previous studies on $TiO_2$ film fabrication have largely focused on photovoltaic, electronic, and thermal applications,[49-51] with limited exploration of the material's anisotropy. In this work, we address this gap and unveil the strong material anisotropy of 2D $TiO_2$. The mild two-step oxidation method we developed enables large-scale fabrication of highly anisotropic 2D $TiO_2$ films. The fabrication of 2D $TiO_2$ from MXene and its derivatives allows mild, solution-based processing without high-temperature treatment required by CVD,[52] and the bottom-up self-assembly also enables precise control over film thickness and large-area coating (with the coating area limited only by substrate and container size). Both are critical for large-scale manufacturing. In addition, our method leverages the intrinsically separated Ti layers in the MXene precursor, facilitating effective separation of $TiO_2$ monolayers and preservation of their structural anisotropy for polarizer applications.

## 4. Material characterization

We employed a range of material characterization techniques to verify the material transformation during the two-step oxidation process and to assess the quality of MXene and its derivatives, including X-ray photoelectron spectroscopy (XPS), Raman spectroscopy, X-ray diffraction (XRD), ultraviolet–visible (UV–vis) absorption spectroscopy, and spectroscopic ellipsometry. For comparison, the same sample, initially consisting of 3 layers of MXene, was characterized at different stages of the two-step oxidation process, including its initial MXene form, the MXO form after the first-step oxidation, and the $TiO_2$ form after the second-step oxidation.

**Figure 3** shows the measured XPS spectra, along with the corresponding fittings in the C 1*s*, Ti 2*p*, and O 1*s* regions. **Figure 3a(i)** shows the C 1*s* region of MXene. Two prominent peaks are observed at ~284.8 eV and ~286.4 eV, indicating the presence of graphitic C–C bonds and C–O bonds. An additional peak at ~281.2 eV corresponds to C–Ti bonds between $Ti_3C_2$ layers. A weak peak at ~288.5 eV, associated with the C–F functional groups, is also observed, likely originating from the LiF/HCl etching process during the synthesis of MXene nanoflakes.[53, 54]

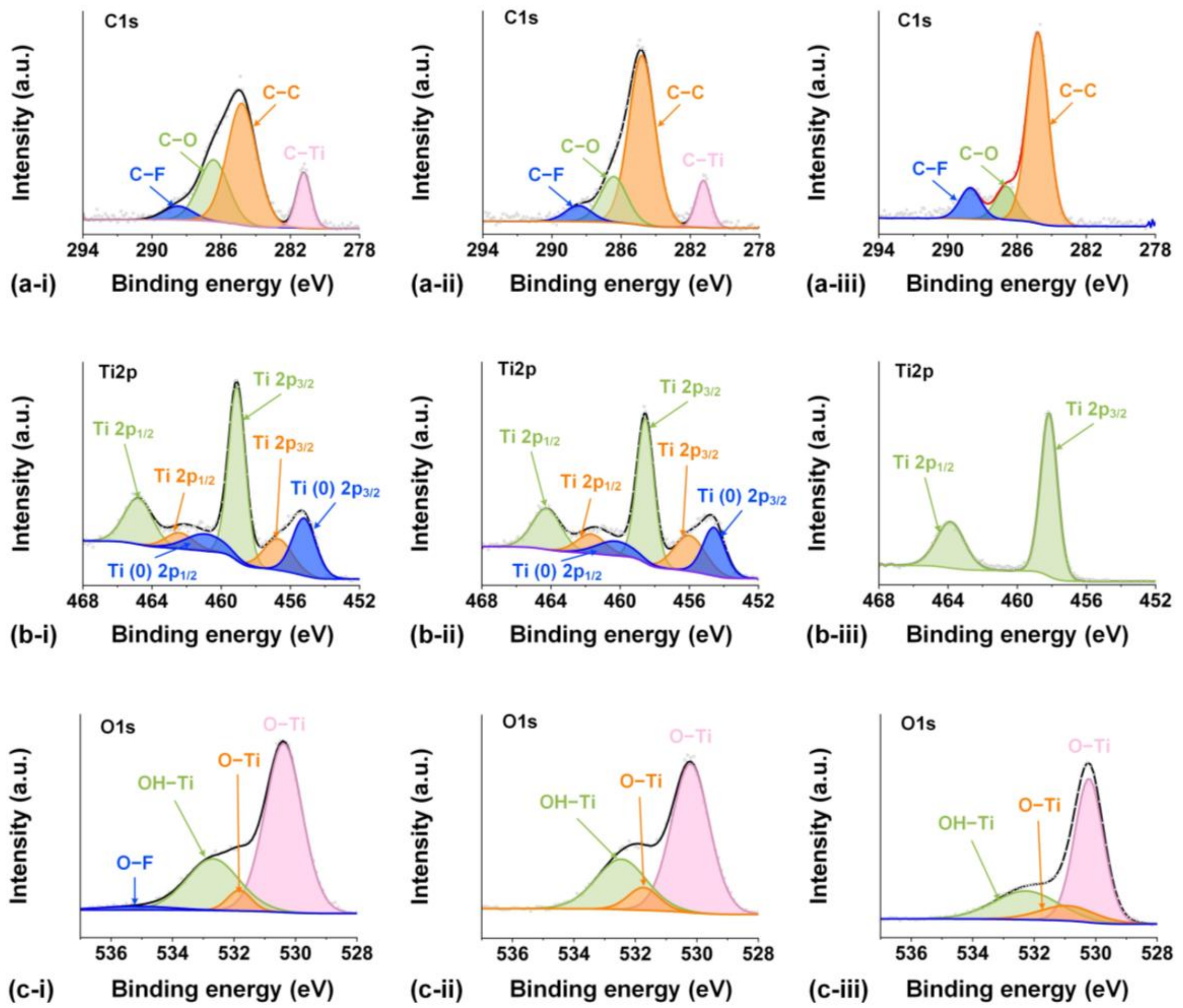

**Figure 3.** X-ray photoelectron spectroscopy (XPS) spectra of (**a**) the C 1*s*, (**b**) the Ti 2*p*, and (**c**) the O 1*s*. In (**a**) – (**c**), (**i**) – (**iii**) show the results for MXene, MXO, and $TiO_2$ during the two-step oxidation process, respectively.

**Figure 3a(ii)** shows the C 1*s* region of MXO. The high-resolution C 1*s* spectrum of MXO is similar to that of MXene, except for a slight decrease in the contribution of Ti–C components with increasing oxidation degree. **Figure 3a(iii)** shows the C 1*s* region of $TiO_2$. The observed C 1*s* signal can be attributed to residual carbon species and adventitious surface carbon.[55] Note that the fitting of C 1*s* region reflects the relative distribution of bonding states within the detected carbon species, rather than the absolute carbon content in the prepared $TiO_2$ film. In addition, the peak at ~281.2 eV, corresponding to C–Ti bonds, disappears. This is attributed to the elimination of C–Ti bonds after the second-step oxidation, consistent with the conversion from a carbide-based MXene to an oxide lattice.[56]

In **Figure 3b**, the Ti 2*p* region is deconvoluted into multiple components corresponding to distinct Ti bonding environments. The area ratio of the Ti $2p_{3/2}$ to Ti $2p_{1/2}$ peaks is fixed at 2:1, in agreement with the value reported in Ref. [57]. **Figure 3b(i)** shows the Ti 2*p* region of MXene. Two peaks associated with C–Ti bonds are observed at ~455.2 eV, and ~460.9 eV, corresponding to the Ti (0) 2*p* doublet of Ti (0) $2p_{3/2}$ and Ti (0) $2p_{1/2}$, respectively. The two peaks at ~456.7 eV and ~462.4 eV are attributed to the $2p_{3/2}$ and $2p_{1/2}$ components of Ti atoms bonded to O/F surface terminations. In addition, the two peaks at ~459.1 eV and ~464.8 eV correspond to another Ti 2*p* doublet of Ti $2p_{3/2}$ and Ti $2p_{1/2}$, respectively. This indicates the presence of Ti–O bonds, resulting from partial surface oxidation or physisorbed oxygen during ambient exposure.

**Figure 3b(ii)** shows the Ti 2*p* region of MXO. The high-resolution Ti 2*p* spectrum of MXO does not exhibit significant differences compared with that of MXene, suggesting that the main Ti–C bonding framework is preserved despite partial surface oxidation. **Figure 3b(iii)** shows the Ti 2*p* region of $TiO_2$. The two prominent peaks at ~458.2 eV and ~463.8 eV are attributed to the Ti 2*p* doublet of Ti $2p_{3/2}$ and Ti $2p_{1/2}$, respectively. This is due to the irreversible breaking of C–Ti bonds with increasing oxidation degree, consistent with the conversion from a carbide-based MXene to an oxide lattice.[56]

**Figure 3c(i)** shows the high-resolution O 1*s* spectrum of MXene. The two prominent peaks at ~530.2 eV and ~531.7 eV are attributed to Ti–O bonds on the surface. The two additional peaks at ~532.5 eV and ~535.2 eV correspond to –OH functional groups bonded to the titanium network and O–F bonds, respectively, likely introduced during the LiF/HCl-etching process. **Figure 3c(ii)** shows the O 1*s* region of MXO. Compared with MXene, the two prominent peaks at ~530.2 eV and ~531.7 eV show an increase in relative intensity. The weak peak at ~535.2 eV, corresponding to O–F bonds, disappears. This is attributed to the replacement of O–F bonds by other oxygen-containing functional groups with increasing oxidation degree.

**Figure 3c(iii)** shows the O 1*s* region of $TiO_2$. Compared with MXene and MXO, the Ti–O contribution becomes stronger, reflecting the formation of a $TiO_2$ lattice. This also indicates that, with increasing oxidation degree, a progressive shift occurs from Ti–C dominated peaks in MXene to Ti–O dominated features in MXO, and ultimately to $Ti^{4+}$ lattice oxygen peaks in $TiO_2$. The trend provides further evidence for a gradual transition from a terminated carbide MXene to a $TiO_2$ phase.

**Figure 4a** shows the Raman spectra, which were measured using a ~514-nm pump laser. For MXene, two narrow peaks at ~203 $cm^{-1}$ and ~734 $cm^{-1}$ are observed, corresponding to the $A_{1g}$ vibrational mode. The weak peak centred at ~374 $cm^{-1}$ corresponds to the in-plane ($E_g$) vibrations influenced by surface functional groups. These results are consistent with previous measurements for MXene in Refs. [58, 59], which confirm the high quality of our prepared MXene film. As the oxidation degree increases, MXO exhibits a slight decrease in the intensity of these peaks, indicating that it undergoes mild oxidation. For $TiO_2$, a characteristic peak at ~140 $cm^{-1}$ is observed, further confirming the formation of $TiO_2$ after the second-step oxidation.[60]

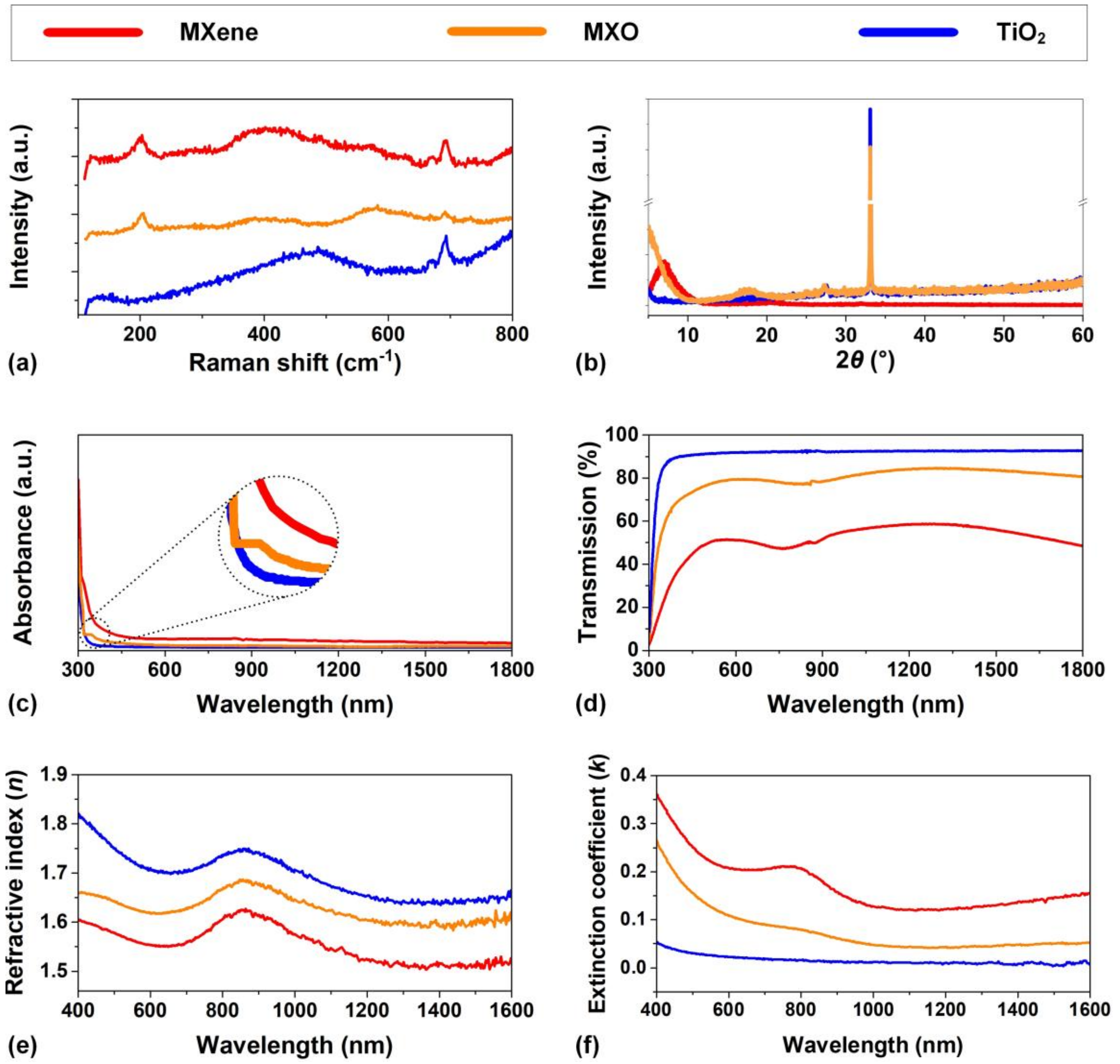

**Figure 4.** Measured (**a**) Raman spectra, (**b**) X-ray diffraction (XRD) spectra, (**c**) ultraviolet-visible (UV–vis) absorption spectra, (**d**) linear transmittance spectra, (**e**) in-plane refractive indices $n$, and (**f**) in-plane extinction coefficients $k$ for MXene, MXO, and $TiO_2$ during the two-step oxidation process.

**Figure 4b** shows the XRD spectra measured using a Bruker D8 X-ray diffractometer operated at 40 kV, with $2\theta$ ranging from 5° to 60°. For MXene, the (002) diffraction peak at $2\theta$ = ~7°, corresponding to the basal planes of the 2D titanium carbide layers, is consistent with that reported in Refs. [61, 62] and further confirms the successful preparation of the MXene film. For MXO, a decrease in the intensity of the peak at $2\theta$ = ~38° is observed, indicating the oxidation of MXene. For $TiO_2$, the XRD peaks do not match with those of anatase and rutile $TiO_2$,[63] suggesting that the $TiO_2$ film is not fully crystallized and contains structural defects, which is likely due to the self-assembly nature of nanoflakes. Despite this, it is worth noting that this does not significantly degrade the performance of the hybrid polarizers, as discussed in **Section 5**. We also infer that these structural defects may contribute to the non-negligible optical absorption observed in the $TiO_2$ films, as crystalline $TiO_2$ should exhibit negligible absorption at infrared wavelengths. A sharp diffraction peak at ~33° is observed for both MXO and $TiO_2$. This peak corresponds to the Si (200) reflection from the silicon substrate, indicating that the oxide layers are sufficiently thin to allow substrate diffraction to dominate in this region.

**Figures 4c** and **d** show the optical absorption and transmittance spectra characterized by UV–vis spectrometry, respectively. Compared with MXene and MXO, $TiO_2$ exhibits lower optical absorption and higher transmittance over the measured wavelength range, which can be attributed to its large optical bandgap. In addition, $TiO_2$ maintains low optical absorption at wavelengths above 400 nm, supporting its application in broadband optical polarizers.

**Figures 4e** and **f** show the refractive indices ($n$) and extinction coefficients ($k$) measured by spectral ellipsometry. Due to the weak out-of-plane response of the 2D films, we could only measure the in-plane $n$ and $k$ values. For MXene, our measured $k$ values are lower than those in Refs. [64, 65], which can be attributed to the differences in the sample size and the surface functional groups. Among the three materials, $TiO_2$ exhibits the highest refractive index and the lowest extinction coefficient, showing agreement with the oxidation trend of MXene reported in Refs. [66, 67].

## 5. Waveguide polarizers

In this section, we characterize the performance of our fabricated waveguide polarizers, including their *PDL*, *FOM*, *OBW*, and power tolerance. For comparison, we perform systematic measurements for hybrid devices with 2D MXene, MXO, and $TiO_2$, including those with various coating lengths ($L_c$) and layer numbers ($N$) of these 2D materials.

**Figure 5** shows the results for *PDL* measurements. In our measurements, lensed fibers were used to butt couple a continuous-wave (CW) laser at ~1550 nm into and out of the fabricated devices with inverse-taper couplers at both ends. The fiber-to-chip coupling loss was ~4 dB / facet. For comparison, all measurements were performed using the same input power of $P_{in}$ = ~0 dBm. Unless otherwise specified, the values of $P_{in}$ and *IL* in our following discussion refer to those after excluding the fiber-to-chip coupling loss. The values of $N$ refer to the numbers of MXene layers, determined by the numbers of repeated layer-by-layer coating cycles in **Figure 2e**. The same values are also used for the corresponding $TiO_2$ layer numbers after the two-step oxidation process. Due to the detection limit of the optical power meter used in our measurements, we do not show the results for *IL* > 70 dB.

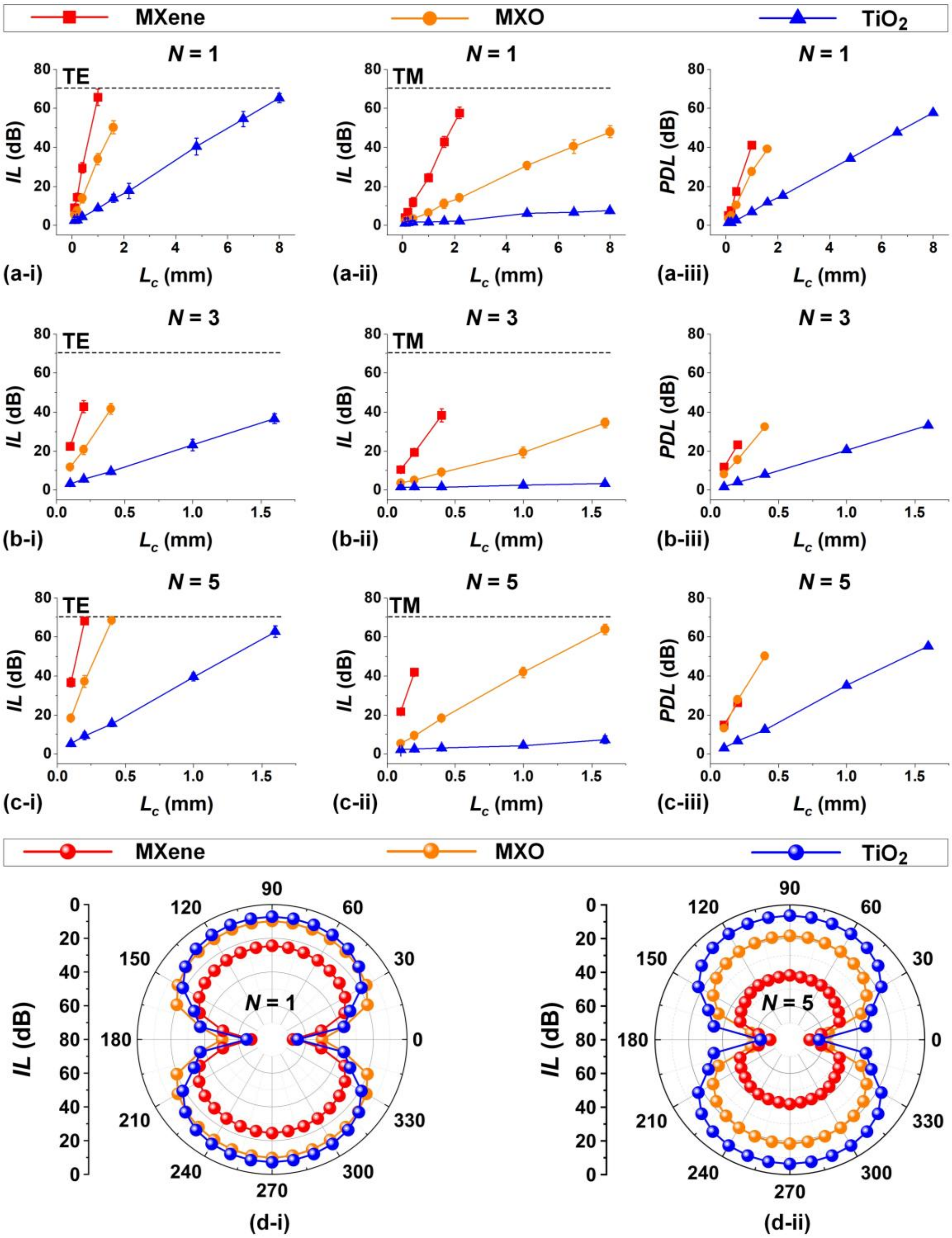

**Figure 5.** Measured insertion loss (*IL*) versus 2D material coating length ($L_c$) for the hybrid waveguides with (**a**) 1, (**b**) 3, and (**c**) 5 layers of MXene, MXO, and $TiO_2$ ($N$ = 1, 3, 5). In (**a**) – (**c**), (**i**) and (**ii**) show the results for TE and TM polarizations, respectively, and (**iii**) shows the polarization-dependent loss (*PDL*) calculated based on the results in (**i**) and (**ii**). The dashed horizontal lines in (**i**) and (**ii**) indicate the maximum detectable *IL* in our experiments. (**d**) Polar diagrams for the measured *IL* of devices with (**i**) 1 and (**ii**) 5 layers of MXene, MXO, and $TiO_2$ ($N$ = 1, 5) that achieve the maximum *PDL* values in (**a-iii**) and (**c-iii**), respectively. The polar angle corresponds to the angle between the input polarization plane and the substrate. In (**a**) – (**d**), the input continuous-wave (CW) power and wavelength are ~0 dBm and ~1550 nm, respectively.

**Figures 5a(i, ii)** show the measured TE- and TM-polarized *IL* versus $L_c$ for the hybrid waveguides with 1 layer of MXene, MXO, and $TiO_2$ ($N$ = 1), respectively. In each figure, the data points depict the average values obtained from the experimental results of three duplicate devices, and the error bars indicate the variations across different devices. **Figure 5a(iii)** shows the *PDL* obtained by subtracting the TM-polarized *IL* in **Figure 5a(ii)** from the TE-polarized *IL* in **Figure 5a(i)**. For all the hybrid waveguides, the *IL* increases with $L_c$ for both TE and TM polarizations, with a more dramatic increase observed for the TE polarization, leading to obvious *PDL* for these devices. We also measured uncoated silicon waveguides and silicon waveguides coated only with a polymer layer (see details in Note S3, Supporting Information), which exhibit negligible *PDL* (< 0.3 dB), confirming that the polarization selectivities of the hybrid waveguides are predominantly determined by the 2D MXene, MXO, and $TiO_2$ films.

For all the hybrid waveguides in **Figure 5a(iii)**, the *PDL* increases with $L_c$, further confirming that high *PDL* can be achieved by increasing the 2D material coating length. We cannot measure the *PDL* for the MXene and MXO devices when $L_c$ > 1.0 mm and $L_c$ > 1.6 mm, respectively. This is because the TE-polarized *IL* (> 70 dB) already exceeds the detection range of the optical power meter. Among the hybrid waveguides, the $TiO_2$ devices show lower *PDL* than MXene and MXO devices with the same $L_c$. This is not surprising, and can be attributed to the relatively low optical absorption of $TiO_2$ for both TE and TM polarizations, which results in a relatively small difference between them. Despite this, the TM-polarized *IL* of the $TiO_2$ devices is very low, yielding a low *EIL* of ~3.3 dB at $L_c$ = 8.0 mm. In contrast, high *EIL* values of > 70 dB and ~42.1 dB are obtained for the MXene and MXO devices with the same $L_c$, respectively. It is worth noting that, although the *PDL* of the $TiO_2$ devices is relatively low at a small $L_c$, it increases with $L_c$ and reaches a high value of ~57.7 dB at $L_c$ = 8.0 mm. At a smaller $L_c$ = 2.2 mm, the *EIL* and *PDL* for the $TiO_2$ device are ~0.9 dB and ~15.7 dB, respectively. The *EIL* is even lower, with the *PDL* still sufficiently large to ensure high polarization selectivity.

**Figures 5b** and **c** show the corresponding results for the hybrid waveguides with 3 and 5 layers of MXene, MXO, and $TiO_2$ ($N$ = 3, 5). For all these devices, the TE-polarized *IL* shows a more obvious increase with $L_c$ than that for TM polarization – a trend similar to that observed in **Figure 5a**. Compared to the devices with $N$ = 1, the devices with thicker 2D films exhibit higher *IL* for both TE and TM polarizations, which is attributed to stronger optical absorption induced by the thicker films. At the same $L_c$, higher *PDL* is achieved for devices with thicker 2D films, reflecting the fact that thicker films enable a faster increase in the *PDL* with increasing $L_c$. For the $TiO_2$ devices with $L_c$ = 1.6 mm, the *PDL* is ~32.6 dB and ~54.4 dB for $N$ = 3 and $N$ = 5, respectively, and the corresponding *EIL* is ~2.6 dB and ~5.0 dB.

In addition to characterizing hybrid waveguides with various coating lengths and 2D material layer numbers, we also characterize hybrid waveguides with different silicon waveguide widths (see Note S4, Supporting Information), finding that a width of 400 nm achieves the best polarizer performance. We also compare $TiO_2$ devices fabricated by one-step and two-step oxidation processes (see details in Note S5,

Supporting Information), with the latter exhibiting better polarizer performance.

**Figures 5d(i, ii)** show the polar diagrams for the measured *IL* of devices with 1 and 5 layers of MXene, MXO, and $TiO_2$ ($N$ = 1, 5). In **Figure 5d(i)**, we show the results for three devices achieving the maximum *PDL* values in **Figure 5a(iii)**, including a MXene device with $L_c$ = ~1.0 mm, an MXO device with $L_c$ = ~1.6 mm, and a $TiO_2$ device with $L_c$ = ~8.0 mm. Similarly, **Figure 5d(ii)** shows the results for the three devices achieving the maximum *PDL* values in **Figure 5c(iii)**. In these polar diagrams, the variations in the *IL* values across different polarization angles further confirm the polarization selectivity of these hybrid waveguides.

In **Figure 6**, we provide further analysis based on the measured results in **Figure 5**. **Figure 6a** shows the calculated polarizer *FOM*s for the MXene, MXO, and $TiO_2$ devices with 1 – 5 layers. The data points depict the average values of devices with different $L_c$, and the error bars indicate the variations among different devices. As can be seen, the $TiO_2$ devices achieve much higher *FOM* values than comparable MXene and MXO devices. A maximum *FOM* value of ~17.5 is achieved for the devices with 1 layer of $TiO_2$. In contrast, the *FOM* values of the 1-layer MXene and MXO devices are only ~1.7 and ~5.1, respectively. In addition, the GO-coated and $MoS_2$-coated silicon waveguides in Refs. [23] and [25] only achieved *FOM* values of ~1.7 (with *PDL* = ~17 dB and *EIL* = ~10 dB) and ~4.2 (with *PDL* = ~21 dB and *EIL* = ~5 dB), respectively. We also note that the *FOM* slightly decreases for increasing layer number $N$. For the $TiO_2$ devices, the *FOM* decreases from ~17.5 to ~10.8 as $N$ increases from 1 to 5. This is a combined result of changes in mode overlap with the 2D films and the unevenness and imperfect interlayer contact in thicker 2D films.

**Figure 6b** shows the waveguide propagation loss (*PL*) versus *N* for the hybrid waveguides with 2D MXene, MXO, and $TiO_2$, which was calculated based on the measured *IL* in **Figures 5a – c**. As expected, the *PL* increases with *N* for both polarizations in these devices, mainly due to the higher light absorption of thicker 2D films. The excess propagation loss (*EPL*) induced by the 2D films was further calculated by excluding the *PL* for the uncoated silicon waveguide, which were ~3.4 dB/cm and ~3.0 dB/cm for TE and TM polarizations, respectively. The TE- and TM-polarized *EPL* induced by monolayer $TiO_2$ (~5-nm-thick) is ~76.2 dB/cm and ~4.1 dB/cm, respectively, in contrast to ~682.0 dB/cm and ~248.7 dB/cm for monolayer MXene (~6-nm-thick), and ~319.4 dB/cm and ~52.7 dB/cm for monolayer MXO (~8-nm-thick). We also note that the value of ~76.2 dB/cm is lower than the typical value of *EPL* induced by monolayer graphene (~2,000 dB/cm,[31] ~0.3-nm-thick) and $MoS_2$ (~117 dB/cm,[25] ~0.7-nm-thick), yet more than three times higher than that induced by monolayer GO (~20 dB/cm,[68] ~2-nm-thick).

**Figure 6c** shows the extinction coefficients of 2D MXene, MXO, and $TiO_2$ for both TE and TM polarizations ($k_{TE}$ and $k_{TM}$), which were obtained by fitting the results in **Figure 6b** with optical mode simulations. For monolayer $TiO_2$, it exhibits a low $k_{TM}$ of ~0.0014, which is ~56 and ~9 times lower than that of monolayer MXene and MXO, respectively. The $k_{TE}$ of monolayer $TiO_2$ (~0.021) is slightly higher than that of monolayer GO (~0.009 [33]), but much lower than those of monolayer graphene (~2.100 [69]) and $MoS_2$ (~0.107 [25]). The non-negligible $k_{TE}$ of $TiO_2$, despite its large bandgap, likely arises from structural defects, as indicated by the XRD results in **Figure 4b**. The slight increase in the extinction coefficients of these 2D materials with layer number *N*, particularly for TE polarization, is likely attributed to increased scattering loss caused by film unevenness, impurity accumulation, and imperfect interlayer contact in thicker films.

In **Figure 6d**, we plot the material anisotropy ratios defined as the ratios of $k_{TE}$ to $k_{TM}$, which reflects the intrinsic anisotropic light absorption of the 2D materials. A maximum anisotropy ratio of ~15 is achieved for monolayer $TiO_2$, which is much higher than ~2 and ~5 for monolayer MXene and MXO, respectively, as well as ~5 and ~4 for monolayer GO and $MoS_2$ reported in Refs. [23] and [25]. This highlights the strong anisotropy in light absorption for 2D $TiO_2$ films.

For 2D-material-coated integrated waveguides, strong anisotropic light absorption arises from both polarization-dependent mode overlap with the lossy 2D materials and their intrinsic loss anisotropy. This is a notable feature for 2D-material-based optical polarizers and also represents a significant advantage over conventional metal-coated waveguide polarizers, in which the polarization selectivity is determined solely by the polarization-dependent mode overlap.[13] In **Figures 6e** and **f**, we analyze the contributions of these two sources to the polarization selectivities of the hybrid waveguides.

In **Figure 6e**, we compare the experimentally measured *PDL* of $TiO_2$-coated waveguides with that obtained via mode simulations using identical *k* values for both TE and TM polarizations (*i.e.*, the *k* values for TE polarization in **Figure 6c**). The simulated values represent the *PDL* induced only by the polarization-dependent mode overlap, and the difference between the measured and simulated *PDL* reflects the contribution from the material loss anisotropy. The negative values for the simulated *PDL* indicate that polarization-dependent mode overlap leads to higher *PL* for TM polarization – in contrast to the measured results, where higher *PL* is observed for TE polarization. In such case, the material loss anisotropy compensates for the negative contribution of the polarization-dependent mode overlap, resulting in higher overall *PL* for TE polarization.

Based on the results in **Figure 6e**, we further plot the fractional contributions ($\eta$'s) of the two sources in **Figure 6f**. The sum of the two fractions is equal to 1, and $\eta$ values $< 0$ and $> 1$ correspond to the case discussed above (*i.e.*, with higher *PL* for TM polarization provided by the polarization-dependent mode overlap). For all layer numbers investigated, the material loss anisotropy contributes over 80% of the overall polarization selectivity, highlighting its dominant role in enabling the functionality of the hybrid polarizers. We also note that the contribution from material loss anisotropy increases with 2D film thickness, mainly due to the decreasing (and eventually negative) contribution from polarization-dependent mode overlap, given that the two contributions always sum to unity.

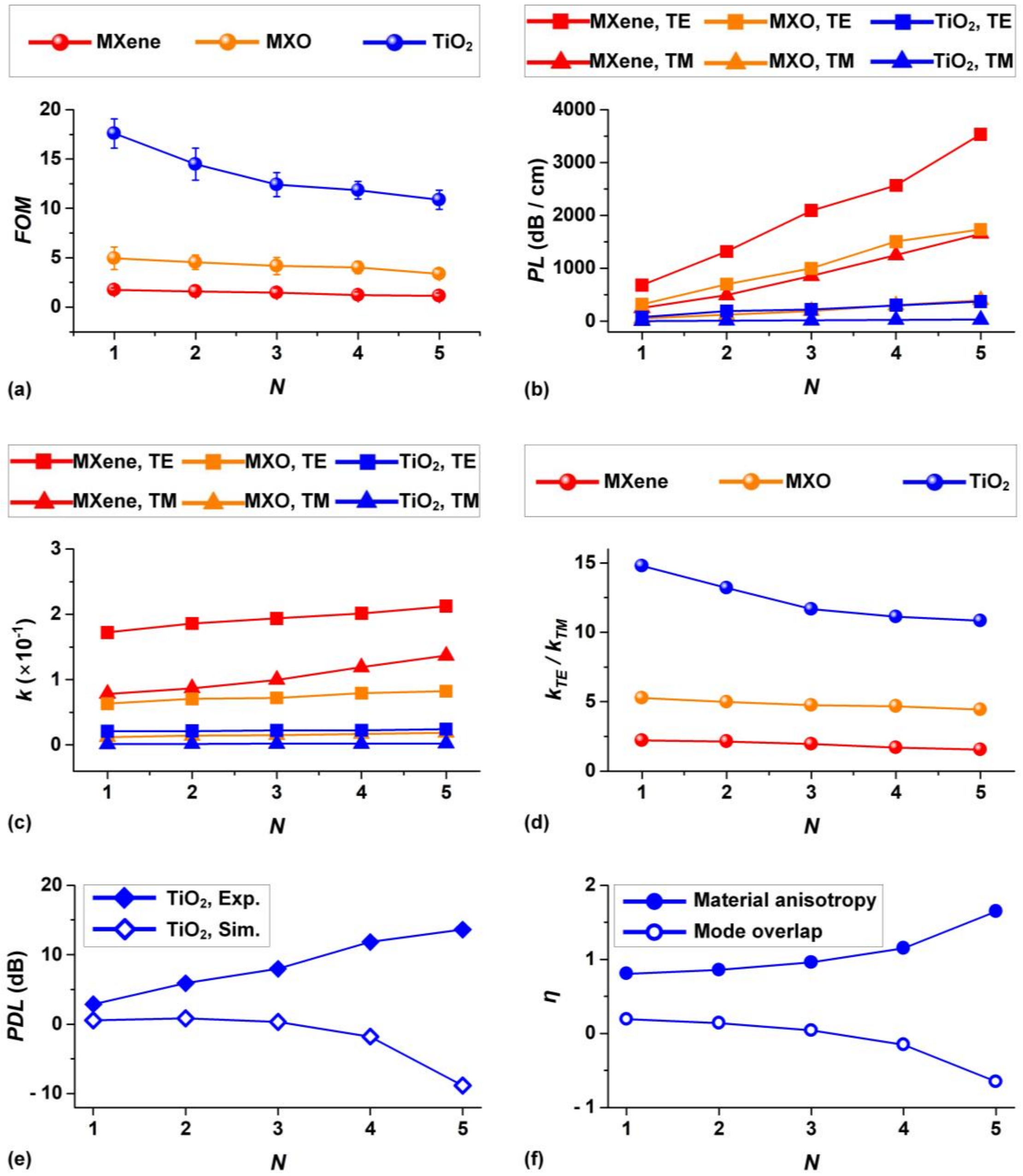

**Figure 6.** (**a**) Polarizer figures of merit (*FOM*) for the waveguide polarizers with 1 – 5 layers of MXene, MXO, and $TiO_2$. The data points depict the average values of devices with different $L_c$, and the error bars indicate the variations among different devices. (**b**) TE- and TM-polarized propagation loss (*PL*) versus layer number *N* for the hybrid waveguides with 2D MXene, MXO, and $TiO_2$. (**c**) TE- and TM-polarized extinction coefficients of 2D MXene, MXO, and $TiO_2$ ($k_{TE}$, $k_{TM}$) versus *N* obtained by fitting the results in (**b**) with optical mode simulations. (**d**) Anisotropy ratio ($k_{TE}$ / $k_{TM}$) calculated from (**c**). (**e**) Measured (Exp.) and simulated (Sim.) *PDL* versus *N* for $TiO_2$ hybrid waveguides with $L_c$ = 0.4 mm. The simulated values, obtained using identical *k* values for both TE and TM polarizations, represent the *PDL* induced solely by polarization-dependent mode overlap. (**f**) Fractional contributions ($\eta$'s) of polarization-dependent mode overlap and material loss anisotropy to the overall *PDL*, which are extracted from (**e**).

For the experiments in **Figure 5**, the *PDL* was measured at a fixed wavelength of

~1550 nm and at a fixed input power of $P_{in}$ = ~0 dBm. In **Figure 7**, we further investigate the dependence of the *PDL* on input CW wavelength and power. **Figure 7a** shows the measured *PDL* versus input CW wavelength for the hybrid waveguides with 1 layer of MXene, MXO, and $TiO_2$ films. Here, we show the results for the three devices in **Figure 5d(i)** that achieve maximum *PDL* values in **Figure 5a(iii)**. **Figure 7b** shows the corresponding results for the $TiO_2$ devices with various layer numbers of $N$ = 1, 3, 5, which achieve the maximum *PDL* values in **Figures 5(a-iii)** – **5(c-iii)**, respectively.

For all the devices in **Figure 7**, no significant changes in the *PDL* (< 2 dB) are observed within the measured wavelength range from ~1500 nm to ~1640 nm, highlighting the broad operation bandwidth for these polarizers, which is usually challenging to achieve for bulk-material-based integrated polarizers.[3, 15] We also notice that there is a minor increase in the *PDL* with increasing wavelength, which can be attributed to dispersion-induced changes in the mode overlap with 2D films. Due to the limited wavelength tuning range of our CW laser, the transmission spectra were measured only within the wavelength range of ~1500 – 1640 nm. According to the results in **Figures 4c** and **4d,** as well as those in previous studies,[70] the response bandwidth of 2D $TiO_2$ can potentially span from the visible to the infrared wavelength region, indicating that the actual *OBW* is expected to extend well beyond the range demonstrated here. In Note S6, Supporting Information, we provide simulation results to estimate the *OBW* of the $TiO_2$ polarizers in **Figure 7b**, showing that the 1-layer device can maintain high *FOM* values (> 10) over a broad wavelength range from 1460 nm to 1800 nm. We also find that there is a trade-off between achieving a high polarizer

*FOM* and a large *OBW* when optimizing the silicon waveguide geometry.

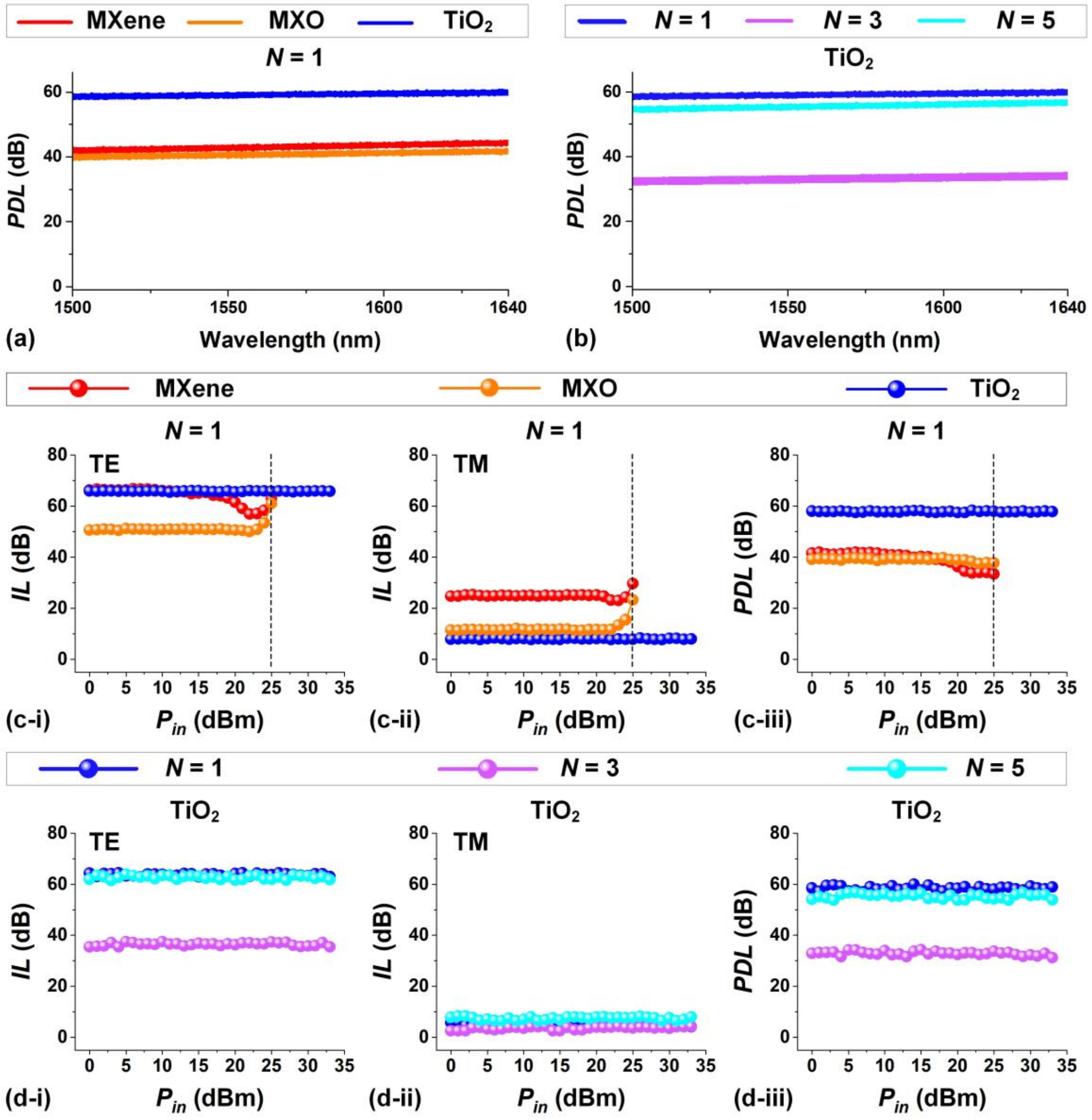

**Figure 7.** (**a**) Measured *PDL* versus wavelength of input continuous-wave (CW) laser for hybrid waveguides with 1 layer of MXene, MXO, and $TiO_2$. (**b**) Measured *PDL* versus input CW wavelength for $TiO_2$ devices with various layer numbers ($N$ = 1, 3, 5). (**c**) – (**d**) Measured (**i**) TE- and (**ii**) TM-polarized *IL* and (**iii**) calculated *PDL* versus input CW power ($P_{in}$) for the devices in (**a**) and (**b**), respectively. In (**c**), the dashed vertical lines indicate the power threshold above which significant thermal damage to the polymer layers occurs. In (**a**) and (**b**), $P_{in}$ = ~0 dBm. In (**c**) and (**d**), the input CW wavelength is ~1550 nm.

**Figures 7c(i, ii)** show the measured TE- and TM-polarized *IL* versus $P_{in}$ for the hybrid waveguides with 1 layer of MXene, MXO, and $TiO_2$. **Figure 7c(iii)** shows the *PDL* calculated based on the results in **Figures 7c(i, ii)**. In our measurements, the input

CW wavelength was maintained at ~1550 nm. The TE-polarized *IL* for the MXene device remained nearly unchanged for $P_{in} \leq$ ~9 dBm, whereas for ~9 dBm $\leq P_{in} \leq$ ~22 dBm, it gradually decreased with increasing $P_{in}$, which can be attributed to the oxidation of MXene in this power range. As $P_{in}$ increased further within ~22 dBm $\leq P_{in} \leq$ ~25 dBm, both the TE- and TM-polarized *IL* started to increase and exhibited a sudden and dramatic rise for $P_{in}$ beyond this range. This is due to the limited thermal stability of the polymer layers in the MXene device introduced during the self-assembly film coating, which cannot withstand the elevated temperatures at such high power levels.[71] The increased *IL* was caused by thermal damage to the polymer layers, which resulted in a permanent decrease in the *PDL* of the MXene device. A similar decrease in the *PDL* was also observed for the MXO device with polymer layers. In contrast, owing to the high thermal stability of $TiO_2$ and the removal of the polymer layers after the piranha treatment, both the TE- and TM-polarized *IL* of the $TiO_2$ device remained nearly unchanged for $P_{in}$ up to ~33 dBm (*i.e.*, ~2 W averaged power coupled into the waveguides), resulting in stable *PDL* at high input powers. It is also worth noting that $P_{in}$ = ~33 dBm was the maximum input power achievable with our experimental setup, corresponding to an initial CW power of ~37 dBm (*i.e.*, ~5 W, amplified using a high-power fiber amplifier) with a 4-dB coupling loss.

**Figure 7d** shows the corresponding results for the $TiO_2$ devices with various layer numbers ($N$ = 1, 3, 5). For these waveguide polarizers with different $TiO_2$ film thicknesses, stable *IL* and *PDL* are achieved within the investigated power range of up to $P_{in}$ = ~33 dBm, further highlighting the excellent high-power tolerance of the $TiO_2$ waveguide polarizers. We also characterized the device temperatures at different $P_{in}$ and analyzed the power tolerance limit (see Note S7, Supporting Information), showing that there remains substantial room to further increase the power within the thermally stable range of the $TiO_2$ films.

To evaluate the stability of the $TiO_2$ polarizers in practical applications, we test their performance under extreme operating conditions. **Figure 8a** shows the testing results under various temperatures, for devices with different $TiO_2$ coating lengths ($L_c$ = ~1.6, ~4.8, ~8.0 mm) and layer numbers ($N$ = 1, 3, 5). Before measuring the *PDL*, the hybrid integrated chips were either cooled in a freezer or heated on a hot plate for 1 h at temperatures ranging from ~−20 °C to 400 °C. As can be seen, the $TiO_2$ polarizers maintain the *PDL* across the investigated temperature range, highlighting their excellent stability at both low and high temperatures, even though the as-prepared $TiO_2$ films are not fully crystallized. We did not further increase the temperature beyond 400 °C, considering the thermal stability limit of the SOI chips and that 400 °C is already sufficient for the majority of practical applications. $TiO_2$ films coated on quartz substrates were also tested after heating at 800 °C or immersing in corrosive chemical environments (see Note S8, Supporting Information), showing negligible changes in the film properties. These results indicate that the $TiO_2$ polarizers can operate in even harsher environments, with potential applications in extreme environments such as volcanoes and outer space. It is also worth mentioning that the above tests were performed for $TiO_2$-coated integrated chips without fiber-pigtail packaging. For packaged devices, the temperature and environmental tolerance is mainly limited by other packaging materials rather than by the $TiO_2$ polarizers themselves. **Figure 8b** shows the testing results for long-term operation, where $TiO_2$ polarizers with different $L_c$ and $N$ were tested at room temperature with $P_{in}$ = ~28 dBm. As can be seen, all the devices maintain stable *PDL* during 48 hours of continuous operation. The same $TiO_2$ polarizers were also tested with $P_{in}$ = ~33 dBm, and they maintain stable *PDL* during 1 hour of continuous operation. We did not continue the test beyond 1 h because the devices were not packaged with fiber pigtails, and the lensed fibers used for light coupling cannot withstand such high power over prolonged operation. These results demonstrate their long-term operational stability for practical applications.

Fabrication yield is also a crucial parameter for industrial deployment. **Figure 8c** shows the testing results for the fabrication yield of devices with 1 layer of $TiO_2$. A total of 100 devices (with different $L_c$) were tested, and an *FOM* value of 15 was set as the decision boundary. Only two devices have *FOM* values below 15, resulting in a high fabrication yield of 98%. The averaged *FOM* across the 98 devices with *FOM* > 15 is ~17.6, and the maximum *FOM* for a single device reaches ~19.3. This reflects the high repeatability of the fabrication process and the consistency in achieving high performance for the fabricated devices.

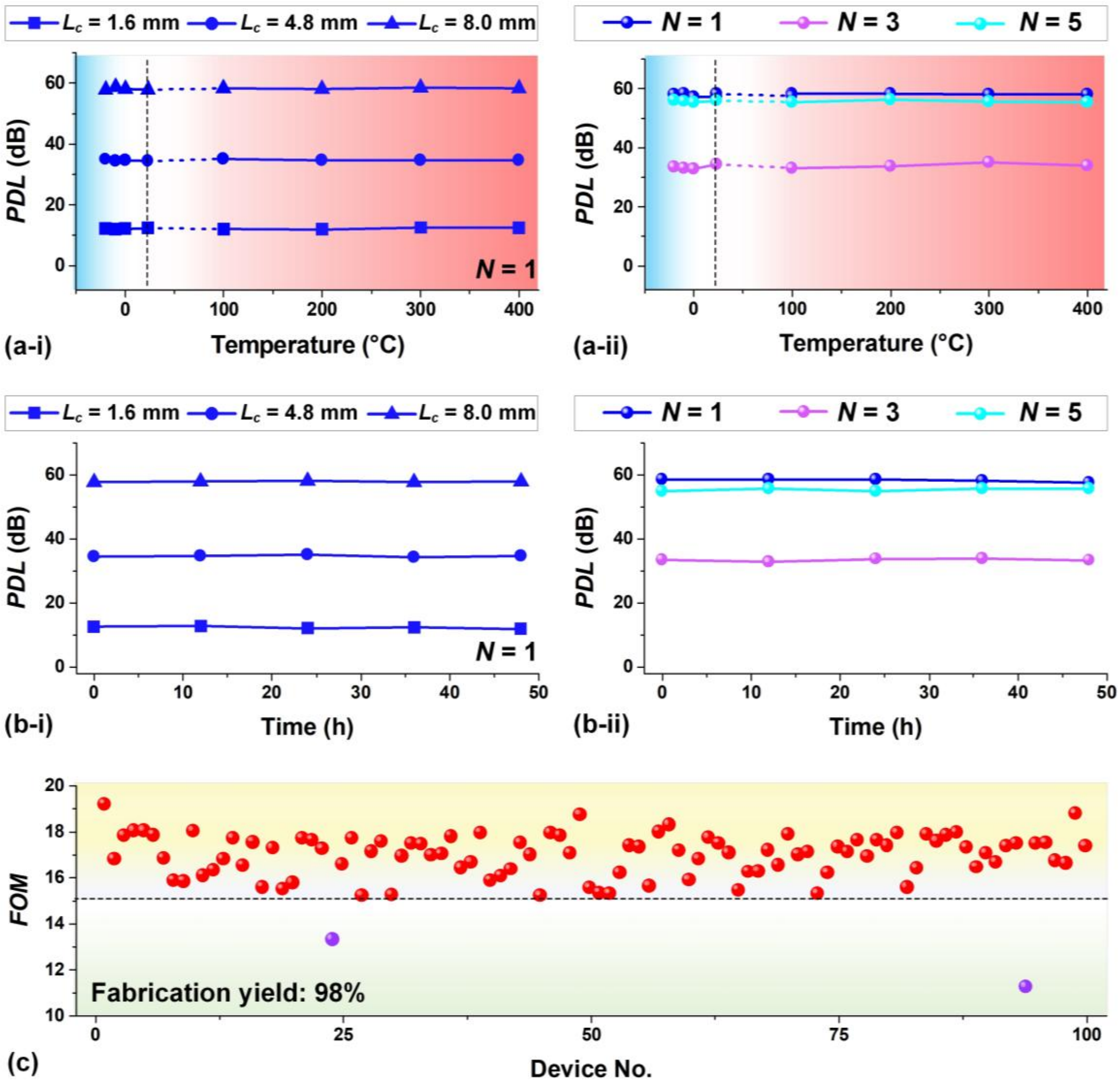

**Figure 8**. (**a**) Measured *PDL* after cooling / heating treatments at various temperatures for $TiO_2$ hybrid waveguide polarizers with (**i**) different $TiO_2$ coating lengths ($L_c$) and (**ii**) different layer numbers ($N$). The hybrid integrated chips were cooled or heated at various temperatures for 1 h prior to the *PDL* measurements. The dashed vertical lines indicate the room temperature without cooling or heating. (**b**) Measured *PDL* versus operation time for $TiO_2$ waveguide polarizers with (**i**) different $L_c$ and (**ii**) different $N$. (**c**) Measured polarizer *FOM*s of 100 waveguide polarizers with 1 layer of $TiO_2$. The dashed line indicates the decision boundary (*FOM* = 15) for calculating the fabrication yield. In (**a**) and (**b**), the input CW power was $P_{in}$ = ~28 dBm.

## 6. Microring resonator polarizers

In addition to waveguide optical polarizers, we also demonstrate microring resonator (MRR) polarizers consisting of silicon MRRs integrated with 2D MXene, MXO, and $TiO_2$. We fabricated silicon MRRs with a radius of ~20 μm, and the 2D film coating length was $L_c$ = ~10 μm. The hybrid MRR and waveguide polarizers were fabricated on the same SOI chip via the same processes.

**Figure 9a** shows a schematic of a silicon MRR integrated with a 2D $TiO_2$ film, and the corresponding microscopic image of the fabricated device is provided in **Figure 9b**. **Figure 9c** shows the TE- and TM-polarized transmission spectra for the MRR polarizers with 1 layer of MXene, MXO, and $TiO_2$. The corresponding results for the uncoated silicon MRR are also shown for comparison. All the spectra were measured at a low input power of ~−10 dBm that did not induce any significant thermo-optic effects. As can be seen, the uncoated silicon MRR does not show significant difference in the extinction ratios (*ER*s) between the TE- and TM-polarized resonances. In contrast, all the hybrid devices exhibit lower *ER*s for the TE polarization. The uncoated silicon MRR is slightly over-coupled,[72] and the integration of 2D films increases the round-trip loss and makes it become under-coupled. Since the in-plane light absorption (for TE polarization) is higher than the out-of-plane absorption (for TM polarization) in the 2D films, the TE polarization suffers higher round-trip loss, which leads to a greater deviation from the critical coupling condition and hence lower *ER*s. Similarly, due to the lower light absorption of $TiO_2$ that enables lower round-trip loss, the $TiO_2$ device exhibits higher *ER*s than the MXene and MXO devices for both TE and TM polarizations.

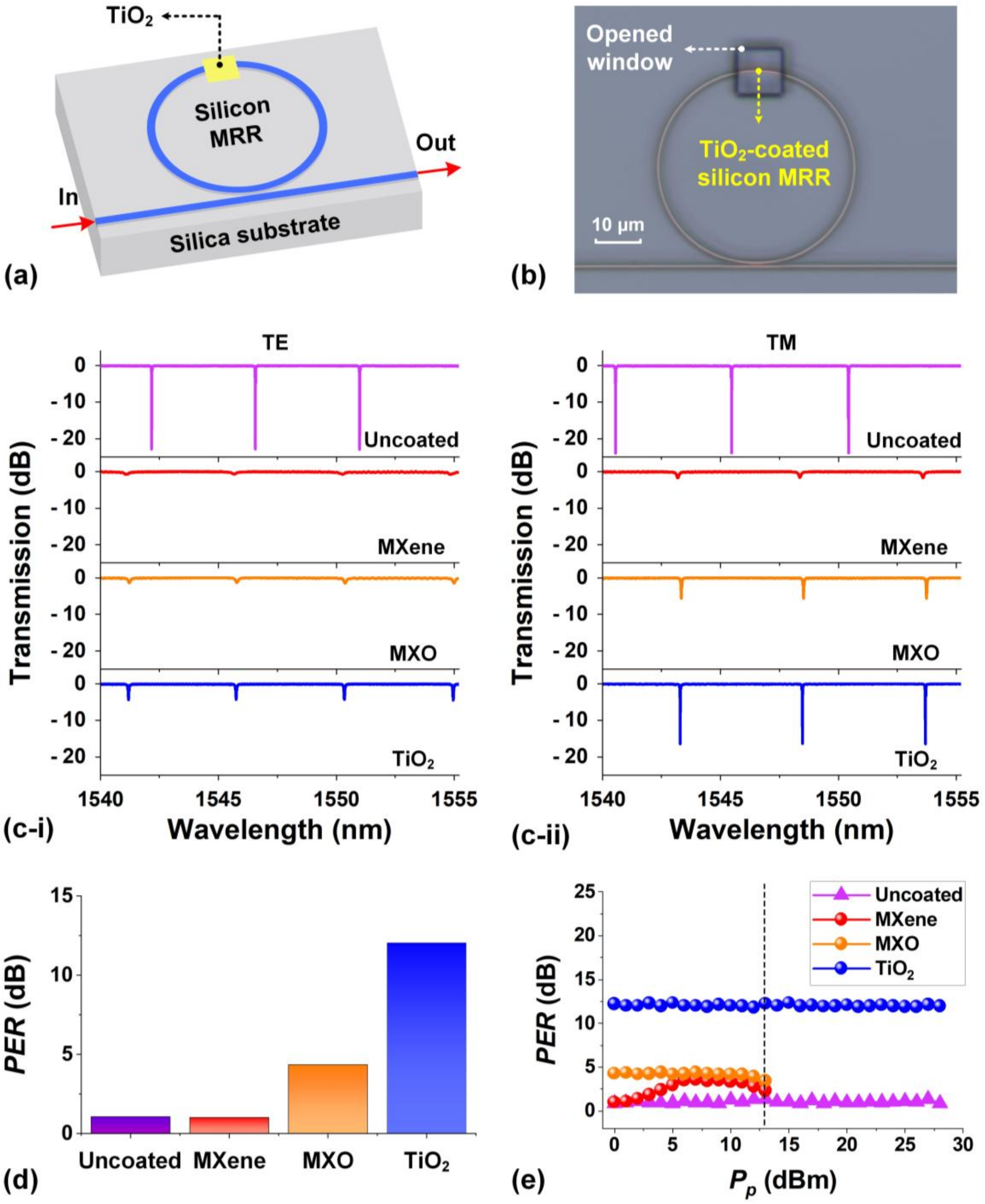

**Figure 9.** (**a**) Schematic of a microring resonator (MRR) polarizer consisting of a silicon MRR integrated with a 2D $TiO_2$ film. (**b**) Microscopic image of a fabricated device with 1 layer of $TiO_2$ ($N$ = 1). (**c**) Measured (**i**) TE- and (**ii**) TM-polarized transmission spectra of MRR polarizers with 1 layer of MXene, MXO, and $TiO_2$. (**d**) Polarization extinction ratios (*PER*'s) extracted from (**c**). In (**c**) – (**d**), the CW input power was ~−10 dBm. (**e**) Measured *PER* versus input CW pump power $P_p$ for hybrid MRRs with 1 layer of MXene, MXO, and $TiO_2$. The dashed vertical line indicates the power threshold above which significant thermal damage to the polymer layers occurs. In (**c**) – (**e**), the corresponding results for the uncoated MRR are also shown for comparison.

**Figure 9d** shows the polarization extinction ratios (*PER*s) of the MRRs in **Figure 9c**. The *PER* is defined as the difference between the TE- and TM-polarized *ER*s. The uncoated silicon MRR has a low *PER* of ~1.0 dB, indicating its negligible polarization selectivity. The $TiO_2$ device achieves a high *PER* of ~12.0 dB, in contrast to ~1.0 dB and ~4.3 dB achieved for the MXene and MXO devices, respectively.

We also characterize the power-dependent response for these MRR polarizers. The results are provided in **Figure 9e**. To mitigate the influence of nonlinear effects (*e.g.*, optical bistability), and improve the accuracy of characterizing the MRR's *ER*s at high powers, we employed a different experimental setup with two CW lasers for pump-probe measurements (see details in Note S9, Supporting Information), including a high-power pump with a power of $P_p$ to pump the MRR and a low-power probe with a power of ~−10 dBm to scan the transmission spectra at various $P_p$.

In **Figure 9e**, the *PER* of the MXene device remains unchanged with increasing $P_p$ at low powers, and begins to increase when $P_p$ > ~2 dBm. This can be attributed to photo-thermally induced oxidation of MXene. It is interesting to note that, unlike the reduction of GO induced by photo-thermal effects, which leads to increased light absorption,[33, 71] the oxidation of MXene shows an opposite trend with decreased light absorption. For the MXene and MXO devices, the *PER* starts to decline when $P_p$ > 10 dBm. This can be attributed to the limited thermal stability of the polymer layers in these devices. As $P_p$ continued to increase beyond 13 dBm, the polymer layers suffered substantial thermal damage, resulting in high loss and a rapid drop in the *PER*. In contrast, the $TiO_2$ device maintains a high *PER* across the entire power range up to $P_p$

= ~28 dBm, which corresponds to an effective intra-cavity power of ~26 dBm given the optical bistability in the hybrid MRR (see details in Note S10, Supporting Information), further highlighting its excellent high-power tolerance. We did not characterize these devices for $P_p$ > ~28 dBm because the high power made it difficult for the MRRs to reach thermal equilibrium during the pump-probe measurements, leading to substantial uncertainty in the measured *ER*s.

## 7. Performance comparison of 2D-material-based optical polarizers

In **Table 1**, we summarize state-of-the-art waveguide optical polarizers incorporating 2D materials and compare their performance. Here we only show the results for experimental works, and compare the key parameters including the 2D film length, *OBW*, *PDL*, *EIL* and *FOM*, as well as high-power tolerance and overall industry readiness.

All the 2D-material-based waveguide polarizers in **Table 1** exhibit broad *OBW*s, representing a key advantage over bulk-material-based polarizers. It is also worth noting that only the *OBW*s demonstrated experimentally are shown here, and the actual *OBW*s could be even broader, since many 2D materials exhibit broadband anisotropic optical responses spanning from the visible to the infrared regions. Most polarizers in **Table 1** achieve high *PDL* values. As discussed previously, this is not challenging and can be realized by increasing the 2D film lengths to accumulate a larger *IL* difference induced by the anisotropic light absorption of 2D materials. However, increasing the 2D film length also results in higher *EIL* and hence greater power consumption, which is undesirable for practical applications. For instance, we note that some polarizers in **Table 1** exhibit high *EIL* values exceeding 6 dB (*i.e.*, power loss > 75 %), which are considerably high for a passive optical device in many practical applications.

The polarizer *FOM* can be used to quantitatively evaluate the trade-off between achieving a high *PDL* and maintaining a low *EIL*, and our polarizers in this work achieve a record high polarizer *FOM* of ~17.5 among all 2D-material-coated devices and also among all devices based on the silicon photonic platform. The length of 2D $TiO_2$ film can also be reduced to achieve a lower *EIL* while maintaining the high *FOM* for demanding applications such as optical communications. In fact, the *FOM* of our polarizers is surpassed only by the ~28.8 reported for a graphene-chalcogenide hybrid waveguide,[5] in which graphene was integrated into the waveguide core to enhance mode overlap, rather than being coated on the waveguide top surface as commonly adopted for other works in **Table 1**. Integrating 2D materials into waveguide cores involves more complex fabrication processes and increased cost, and therefore has not been widely employed for 2D-material-based optical polarizers. Compared with the chalcogenide platform,[73] the silicon photonic platform for our polarizers is more mature and widely used.[74] Developing 2D-material-based integrated polarizers based on the silicon photonic platform can not only leverage CMOS fabrication technologies with well-developed process design kits (PDKs) to facilitate large-scale manufacturing,[36] but also enable strong compatibility with existing integrated optical systems (consisting of diverse modules such as modulators, detectors, and electronic-photonic co-integration),[75, 76] thereby greatly expanding their application scope.

In addition to the high polarizer *FOM*, our polarizers also exhibit excellent tolerance to high optical powers, making them robust for practical applications, particularly in high-power systems. In contrast, many polarizers in **Table 1**, such as those based on GO and polymers, show limited high-power tolerance.

Finally, the fabrication process of our polarizers, including transfer-free, layer-by-layer film coating with precise control, in situ material transformation, and the fabrication of silicon devices using CMOS technologies, demonstrates high compatibility with integrated device manufacturing and holds strong potential for industrial-scale production. The capabilities for operating in extreme environments, the long-term operational stability, and the high fabrication yield demonstrated in this work further highlight the practical value and the industrial development potential of our polarizers. These, together with the outstanding overall performance metrics, enable our polarizers to achieve a high level of readiness for industrial deployment. This work has broad implications for microcombs [84-114] and their applications to microwave photonics, neuromorphic processors and communications. [115 – 170] The addition and use of 2D materials [171 – 212] will add extra functionality to microcomb chips for potential applications to quantum photonics [213 – 229] and other areas. [230 – 247]

**Table 1. Comparison of waveguide optical polarizers incorporating 2D materials.** WG: waveguide, *OBW*: operational bandwidth, *PDL*: polarization-dependent loss, *EIL*: excess insertion loss induced by 2D materials, *FOM*: figure of merit.

| 2D / WG materials | *OBW* (μm)[a] | 2D film length (mm) | *PDL* (dB) | *EIL* (dB) | *FOM* | High power tolerance | Industry readiness | Ref. |
|---|---|---|---|---|---|---|---|---|
| graphene / polymer | –[b] | ~7.0 | ~19.8 | ~26.0 | ~0.8 | No[c] | low | [77] |
| graphene / glass | ~1.23 –1.61 | ~4.0 | ~27.3 | ~9.0 | ~3.0 | – | low | [78] |
| graphene / $AlF_3$ glass | ~3.15 –3.30 | ~10.0 | ~18.7 | ~1.5 | ~12.2 | – | moderate | [79] |
| graphene / chalcogenide | ~0.94 –1.60 | ~0.4 | ~23.0 | ~0.8 | ~28.8[d] | – | moderate | [5] |
| graphene / polymer | – | – | ~6.4 | ~9.0 | ~0.7 | No | low | [17] |
| $MoSe_2$ / polymer | ~0.98 –1.55 | ~1.2 | ~14.0 | – | – | No | low | [80] |
| $MoS_2$ / Nd:YAG | – | ~10.0 | ~3.0 | ~0.4 | ~7.5 | – | moderate | [81] |
| $MoS_2$ / polymer | ~0.65 –0.98 | – | ~12.6 | – | – | No | low | [82] |
| $MoS_2$ / silicon | ~1.50 –1.60 | ~2.2 | ~21.0 | ~5.0 | ~4.2 | – | moderate | [25] |
| GO / polymer | ~1.53 –1.63 | ~1.3 | ~40.0 | ~6.5 | ~6.2 | No | moderate | [83] |
| GO / doped silica | ~0.63 –1.60 | ~2.0 | ~54.0 | ~7.5 | ~7.2 | No | moderate | [16] |
| GO / silicon | ~1.50 –1.60 | ~2.2 | ~17.0 | ~10.3 | ~1.7 | No | low | [23] |
| rGO / silicon | ~1.50 –1.60 | ~0.4 | ~47.0 | ~15.7 | ~3.0 | No | low | [24] |
| $TiO_2$ / silicon | ~1.50 –1.64 | ~2.2 ~8.0 | ~15.7 ~57.7 | ~0.9 ~3.3 | ~17.5[e] | Yes | high | This work |

[a] Here we only show the *OBW* demonstrated in experiments. [b] No reported value or experimental demonstration in the literature. [c] 'Yes' and 'No' indicate whether the device can maintain its performance at an input power of 2 W. [d] The high FOM was achieved by integrating graphene into the waveguide core to enhance mode overlap, rather than coating 2D materials on the waveguide top surface as commonly adopted in other works. [e] This value is averaged for different devices, and the maximum value for a single device is ~19.3.

## 8. Conclusion

In summary, we experimentally demonstrate 2D-material-based integrated optical polarizers with a high level of industry readiness by integrating 2D MXene films onto silicon photonic devices and subsequently converting them in-situ into $TiO_2$. Compared with MXene, 2D $TiO_2$ exhibits significantly improved anisotropic light absorption and thermal / chemical stability. Experimental results show that the hybrid integrated polarizers achieve a record high polarizer *FOM* of ~17.5 (among 2D-material-coated devices), excellent power tolerance (stable under >2 W coupled average power), a broad operation bandwidth (> 140 nm), and a high fabrication yield (~98%). We also demonstrate $TiO_2$-coated MRR polarizers achieving high *PER*s up to ~12 dB. With combined performance metrics and strong capability for large-scale manufacturing, these 2D-material-based polarizers pave the way for industry-ready implementations and future commercial deployment.

## Conflict of interest

The authors declare no competing financial interest.

## References

[1] D. Dai, J. Bauters, and J. E. Bowers, "Passive technologies for future large-scale photonic integrated circuits on silicon: polarization handling, light non-reciprocity and loss reduction," *Light: Science & Applications,* vol. 1, no. 3, pp. e1-e1, 2012/03/01, 2012.

[2] R. Chipman, W. S. T. Lam, and G. Young, *Polarized light and optical systems*: CRC press, 2018.

[3] D. Dai, L. Liu, S. Gao, D.-X. Xu, and S. He, "Polarization management for silicon photonic integrated circuits," *Laser & Photonics Reviews,* vol. 7, no. 3, pp. 303-328, 2013.

[4] Q. Bao, H. Zhang, B. Wang, Z. Ni, C. H. Y. X. Lim, Y. Wang, D. Y. Tang, and K. P. Loh, "Broadband graphene polarizer," *Nature Photonics,* vol. 5, no. 7, pp. 411-415, 2011/07/01, 2011.

[5] H. Lin, Y. Song, Y. Huang, D. Kita, S. Deckoff-Jones, K. Wang, L. Li, J. Li, H. Zheng, Z. Luo, H. Wang, S. Novak, A. Yadav, C.-C. Huang, R.-J. Shiue, D. Englund, T. Gu, D. Hewak, K. Richardson,

J. Kong, and J. Hu, "Chalcogenide glass-on-graphene photonics," *Nature Photonics,* vol. 11, no. 12, pp. 798-805, 2017/12/01, 2017.

[6] F. Zhang, X. Chen, M. Zhang, W. Jin, S. Han, Z. Yang, and S. Pan, "An excellent deep-ultraviolet birefringent material based on [BO2]∞ infinite chains," *Light: Science & Applications,* vol. 11, no. 1, pp. 252, 2022/08/12, 2022.

[7] X. Chen, B. Zhang, F. Zhang, Y. Wang, M. Zhang, Z. Yang, K. R. Poeppelmeier, and S. Pan, "Designing an Excellent Deep-Ultraviolet Birefringent Material for Light Polarization," *Journal of the American Chemical Society,* vol. 140, no. 47, pp. 16311-16319, 2018/11/28, 2018.

[8] K. Serkowski, D. S. Mathewson, and V. L. Ford, "Wavelength dependence of interstellar polarization and ratio of total to selective extinction," *The Astrophysical Journal,* vol. 196, pp. 261-290, February 01, 1975, 1975.

[9] R. Appel, C. D. Dyer, and J. N. Lockwood, "Design of a broadband UV-visible α-barium borate polarizer," *Applied Optics,* vol. 41, no. 13, pp. 2470-2480, 2002/05/01, 2002.

[10] J.-l. Kou, J.-h. Chen, Y. Chen, F. Xu, and Y.-q. Lu, "Platform for enhanced light–graphene interaction length and miniaturizing fiber stereo devices," *Optica,* vol. 1, no. 5, pp. 307-310, 2014/11/20, 2014.

[11] D. Dai, Z. Wang, N. Julian, and J. E. Bowers, "Compact broadband polarizer based on shallowly-etched silicon-on-insulator ridge optical waveguides," *Optics Express,* vol. 18, no. 26, pp. 27404-27415, 2010/12/20, 2010.

[12] Y. Huang, S. Zhu, H. Zhang, T.-Y. Liow, and G.-Q. Lo, "CMOS compatible horizontal nanoplasmonic slot waveguides TE-pass polarizer on silicon-on-insulator platform," *Optics Express,* vol. 21, no. 10, pp. 12790-12796, 2013/05/20, 2013.

[13] H. Zafar, and M. F. Pereira, "Recent Progress in Light Polarization Control Schemes for Silicon Integrated Photonics," *Laser & Photonics Reviews,* vol. 18, no. 11, pp. 2301025, 2024.

[14] Z. Li, B. Xu, D. Liang, and A. Pan, "Polarization-Dependent Optical Properties and Optoelectronic Devices of 2D Materials," *Research,* vol. 2020, 2020.

[15] Y. Zhang, J. Wu, L. Jia, D. Jin, B. Jia, X. Hu, D. Moss, and Q. Gong, "Advanced optical polarizers based on 2D materials," *npj Nanophotonics,* vol. 1, no. 1, pp. 28, 2024/07/17, 2024.

[16] J. Wu, Y. Yang, Y. Qu, X. Xu, Y. Liang, S. T. Chu, B. E. Little, R. Morandotti, B. Jia, and D. J. Moss, "Graphene Oxide Waveguide and Micro-Ring Resonator Polarizers," *Laser & Photonics Reviews,* vol. 13, no. 9, pp. 1900056, 2019.

[17] J. T. Kim, and H. Choi, "Polarization Control in Graphene-Based Polymer Waveguide Polarizer," *Laser & Photonics Reviews,* vol. 12, no. 10, pp. 1800142, 2018.

[18] Y. Hu, D. Liang, K. Mukherjee, Y. Li, C. Zhang, G. Kurczveil, X. Huang, and R. G. Beausoleil, "III/V-on-Si MQW lasers by using a novel photonic integration method of regrowth on a bonding template," *Light: Science & Applications,* vol. 8, no. 1, pp. 93, 2019/10/09, 2019.

[19] Y. A. Bioud, A. Boucherif, M. Myronov, A. Soltani, G. Patriarche, N. Braidy, M. Jellite, D. Drouin, and R. Arès, "Uprooting defects to enable high-performance III–V optoelectronic devices on silicon," *Nature Communications,* vol. 10, no. 1, pp. 4322, 2019/09/20, 2019.

[20] A. K. Geim, and I. V. Grigorieva, "Van der Waals heterostructures," *Nature,* vol. 499, no. 7459, pp.

419-425, 2013/07/01, 2013.

[21] H. Rokni, and W. Lu, “Direct measurements of interfacial adhesion in 2D materials and van der Waals heterostructures in ambient air,” *Nature Communications,* vol. 11, no. 1, pp. 5607, 2020/11/05, 2020.

[22] B. Radisavljevic, A. Radenovic, J. Brivio, V. Giacometti, and A. Kis, “Single-layer MoS2 transistors,” *Nature Nanotechnology,* vol. 6, no. 3, pp. 147-150, 2011/03/01, 2011.

[23] D. Jin, J. Wu, J. Hu, W. Liu, Y. Zhang, Y. Yang, L. Jia, D. Huang, B. Jia, and D. J. Moss, “Silicon photonic waveguide and microring resonator polarizers incorporating 2D graphene oxide films,” *Applied Physics Letters,* vol. 125, no. 5, 2024.

[24] J. Hu, J. Wu, D. Jin, W. Liu, Y. Zhang, Y. Yang, L. Jia, Y. Wang, D. Huang, B. Jia, and D. J. Moss, “Integrated photonic polarizers with 2D reduced graphene oxide,” *Opto-Electronic Science,* vol. 4, no. 5, pp. 240032-1-240032-16, 2024-12-10, 2025.

[25] J. Hu, J. Wu, I. H. Abidi, D. Jin, Y. Zhang, J. Mao, A. Pandey, Y. Wang, S. Walia, and D. J. Moss, “Silicon Photonic Waveguide Polarizers Integrated With 2D MoS2 Films,” *IEEE Journal of Selected Topics in Quantum Electronics,* vol. 32, no. 2: 3-D Horizons in Photonics: Integrated Circuits, pp. 1-11, 2025.

[26] M. Naguib, M. Kurtoglu, V. Presser, J. Lu, J. Niu, M. Heon, L. Hultman, Y. Gogotsi, and M. W. Barsoum, “Two-Dimensional Nanocrystals Produced by Exfoliation of Ti3AlC2,” *Advanced Materials,* vol. 23, no. 37, pp. 4248-4253, 2011.

[27] S. L. Murray, S. Serajian, S. I. Gnani Peer Mohamed, S. Robinson, R. Krishnamoorthy, S. R. Das, M. Bavarian, S. Nejati, U. Kilic, M. Schubert, and M. Ghashami, “Ultrabroadband Optical Properties of 2D Titanium Carbide MXene,” *ACS Applied Materials & Interfaces,* vol. 16, no. 51, pp. 70763-70773, 2024/12/25, 2024.

[28] J. Schmitz, “Low temperature thin films for next-generation microelectronics (invited),” *Surface and Coatings Technology,* vol. 343, pp. 83-88, 2018/06/15/, 2018.

[29] L. Sirghi, O. Kylián, D. Gilliland, G. Ceccone, and F. Rossi, “Cleaning and Hydrophilization of Atomic Force Microscopy Silicon Probes,” *The Journal of Physical Chemistry B,* vol. 110, no. 51, pp. 25975-25981, 2006/12/01, 2006.

[30] M. Murtaza, S. Sajjad, T. Bano, M. D. K. Leghari, and A. Shahid Malik, “MXene-derived Ti3C2–TiO2/BiOCl heterojunctions with enhanced visible-light photocatalysis and electrochemical activity,” *Nanoscale*, 2026.

[31] H. Cai, Y. Cheng, H. Zhang, Q. Huang, J. Xia, R. Barille, and Y. Wang, “Enhanced linear absorption coefficient of in-plane monolayer graphene on a silicon microring resonator,” *Optics Express,* vol. 24, no. 21, pp. 24105-24116, 2016/10/17, 2016.

[32] J. Li, L. Liu, and T.-K. Sham, “2D XANES–XEOL Spectroscopy Studies of Morphology-Dependent Phase Transformation and Corresponding Luminescence from Hierarchical TiO2 Nanostructures,” *Chemistry of Materials,* vol. 27, no. 8, pp. 3021-3029, 2015/04/28, 2015.

[33] J. Hu, J. Wu, W. Liu, D. Jin, H. E. Dirani, S. Kerdiles, C. Sciancalepore, P. Demongodin, C. Grillet, C. Monat, D. Huang, B. Jia, and D. J. Moss, “2D graphene oxide: a versatile thermo-optic material,” *Advanced Functional Materials,* vol. 34, no. 46, pp. 2406799, 2024.

[34] S. Park, A. T. Garcia-Esparza, H. Abroshan, B. Abraham, J. Vinson, A. Gallo, D. Nordlund, J. Park, T. R. Kim, L. Vallez, R. Alonso-Mori, D. Sokaras, and X. Zheng, "Operando Study of Thermal Oxidation of Monolayer MoS2," *Advanced Science,* vol. 8, no. 9, pp. 2002768, 2021.

[35] L. Liu, S. Ryu, M. R. Tomasik, E. Stolyarova, N. Jung, M. S. Hybertsen, M. L. Steigerwald, L. E. Brus, and G. W. Flynn, "Graphene Oxidation: Thickness-Dependent Etching and Strong Chemical Doping," *Nano Letters,* vol. 8, no. 7, pp. 1965-1970, 2008/07/01, 2008.

[36] W. Bogaerts, D. Pérez, J. Capmany, D. A. B. Miller, J. Poon, D. Englund, F. Morichetti, and A. Melloni, "Programmable photonic circuits," *Nature,* vol. 586, no. 7828, pp. 207-216, 2020/10/01, 2020.

[37] Q. Xu, B. Schmidt, S. Pradhan, and M. Lipson, "Micrometre-scale silicon electro-optic modulator," *Nature,* vol. 435, no. 7040, pp. 325-327, 2005/05/01, 2005.

[38] M. Ghidiu, M. R. Lukatskaya, M.-Q. Zhao, Y. Gogotsi, and M. W. Barsoum, "Conductive two-dimensional titanium carbide 'clay' with high volumetric capacitance," *Nature,* vol. 516, no. 7529, pp. 78-81, 2014/12/01, 2014.

[39] Y.-K. Peng, Y. Hu, H.-L. Chou, Y. Fu, I. F. Teixeira, L. Zhang, H. He, and S. C. E. Tsang, "Mapping surface-modified titania nanoparticles with implications for activity and facet control," *Nature Communications,* vol. 8, no. 1, pp. 675, 2017/09/22, 2017.

[40] M. Dahl, Y. Liu, and Y. Yin, "Composite Titanium Dioxide Nanomaterials," *Chemical Reviews,* vol. 114, no. 19, pp. 9853-9889, 2014/10/08, 2014.

[41] A. Nilsson, L. Pettersson, and J. K. Nørskov, *Chemical Bonding at Surfaces and Interfaces*, 2008.

[42] L. Jia, J. Wu, Y. Zhang, Y. Qu, B. Jia, Z. Chen, and D. J. Moss, "Fabrication Technologies for the On-Chip Integration of 2D Materials," *Small Methods,* vol. 6, no. 3, pp. 2101435, 2022.

[43] Y. Zhang, J. Wu, L. Jia, Y. Qu, Y. Yang, B. Jia, and D. J. Moss, "Graphene Oxide for Nonlinear Integrated Photonics," *Laser & Photonics Reviews,* vol. 17, no. 3, pp. 2200512, 2023/03/01, 2023.

[44] J. Wu, H. Lin, D. J. Moss, K. P. Loh, and B. Jia, "Graphene oxide for photonics, electronics and optoelectronics," *Nature Reviews Chemistry,* vol. 7, no. 3, pp. 162-183, 2023/03/01, 2023.

[45] J. Wu, L. Jia, Y. Zhang, Y. Qu, B. Jia, and D. J. Moss, "Graphene Oxide for Integrated Photonics and Flat Optics," *Advanced Materials,* vol. 33, no. 3, pp. 2006415, 2021.

[46] Y. Zhang, J. Wu, J. Hu, L. Jia, D. Jin, B. Jia, X. Hu, D. J. Moss, and Q. Gong, "2D material integrated photonics: Toward industrial manufacturing and commercialization," *APL Photonics,* vol. 10, no. 4, 2025.

[47] J. Song, M. Zheng, B. Zhang, Q. Li, F. Wang, L. Ma, Y. Li, C. Zhu, L. Ma, and W. Shen, "Fast Growth of Highly Ordered TiO2 Nanotube Arrays on Si Substrate under High-Field Anodization," *Nano-Micro Letters,* vol. 9, no. 2, pp. 13, 2016/11/09, 2016.

[48] A. Khan, S. M. Islam, S. Ahmed, R. R. Kumar, M. R. Habib, K. Huang, M. Hu, X. Yu, and D. Yang, "Direct CVD Growth of Graphene on Technologically Important Dielectric and Semiconducting Substrates," *Advanced Science,* vol. 5, no. 11, pp. 1800050, 2018.

[49] Y. Dong, M. Abbasi, J. Meng, L. German, C. Carlos, J. Li, Z. Zhang, D. Morgan, J. Hwang, and X. Wang, "Substantial lifetime enhancement for Si-based photoanodes enabled by amorphous TiO2 coating with improved stoichiometry," *Nature Communications,* vol. 14, no. 1, pp. 1865,

2023/04/04, 2023.

[50] A. Grishin, B. Bérini, M. Vallet, S. Hurand, F. Maudet, C. Sartel, M. Frégnaux, S. Nowak, G. Amiri, S. Hassani, D. Aureau, V. Sallet, V. Demange, and Y. Dumont, "Tailoring crystallisation of anatase TiO2 ultra-thin films grown by atomic layer deposition using 2D oxides as growth template," *Applied Surface Science,* vol. 641, pp. 158446, 2023/12/30/, 2023.

[51] R. Li, X. Ma, J. Li, J. Cao, H. Gao, T. Li, X. Zhang, L. Wang, Q. Zhang, G. Wang, C. Hou, Y. Li, T. Palacios, Y. Lin, H. Wang, and X. Ling, "Flexible and high-performance electrochromic devices enabled by self-assembled 2D TiO2/MXene heterostructures," *Nature Communications,* vol. 12, no. 1, pp. 1587, 2021/03/11, 2021.

[52] D. Byun, Y. Jin, B. Kim, J. Kee Lee, and D. Park, "Photocatalytic TiO2 deposition by chemical vapor deposition," *Journal of Hazardous Materials,* vol. 73, no. 2, pp. 199-206, 2000/04/03/, 2000.

[53] X. Zhan, C. Si, J. Zhou, and Z. Sun, "MXene and MXene-based composites: synthesis, properties and environment-related applications," *Nanoscale Horizons,* vol. 5, no. 2, pp. 235-258, 2020.

[54] T. Zhang, L. Pan, H. Tang, F. Du, Y. Guo, T. Qiu, and J. Yang, "Synthesis of two-dimensional Ti3C2Tx MXene using HCl+LiF etchant: Enhanced exfoliation and delamination," *Journal of Alloys and Compounds,* vol. 695, pp. 818-826, 2017/02/25/, 2017.

[55] B. M. Hunter, W. Hieringer, J. R. Winkler, H. B. Gray, and A. M. Müller, "Effect of interlayer anions on [NiFe]-LDH nanosheet water oxidation activity," *Energy & Environmental Science,* vol. 9, no. 5, pp. 1734-1743, 2016.

[56] R. A. Soomro, P. Zhang, B. Fan, Y. Wei, and B. Xu, "Progression in the Oxidation Stability of MXenes," *Nano-Micro Letters,* vol. 15, no. 1, pp. 108, 2023/04/18, 2023.

[57] Z. Dessoliers, A. Chemin, G. Valurouthu, R. Lord, T. Bilyk, Y. Gogotsi, V. Mauchamp, and T. Petit, "Combining X-Ray Photoelectron and Absorption Spectroscopies for Determining Surface Chemistry and Composition of Ti3C2Tx MXene," *Advanced Materials Interfaces,* vol. 12, no. 13, pp. 2500391, 2025/07/01, 2025.

[58] A. Sarycheva, M. Shanmugasundaram, A. Krayev, and Y. Gogotsi, "Tip-Enhanced Raman Scattering Imaging of Single- to Few-Layer Ti3C2Tx MXene," *ACS Nano,* vol. 16, no. 4, pp. 6858-6865, 2022/04/26, 2022.

[59] Q. Zhang, H. Lai, R. Fan, P. Ji, X. Fu, and H. Li, "High Concentration of Ti3C2Tx MXene in Organic Solvent," *ACS Nano,* vol. 15, no. 3, pp. 5249-5262, 2021/03/23, 2021.

[60] W. T. Pawlewicz, G. J. Exarhos, and W. E. Conaway, "Structural characterization of TiO2 optical coatings by Raman spectroscopy," *Applied Optics,* vol. 22, no. 12, pp. 1837-1840, 1983/06/15, 1983.

[61] H.-J. Koh, S. J. Kim, K. Maleski, S.-Y. Cho, Y.-J. Kim, C. W. Ahn, Y. Gogotsi, and H.-T. Jung, "Enhanced Selectivity of MXene Gas Sensors through Metal Ion Intercalation: In Situ X-ray Diffraction Study," *ACS Sensors,* vol. 4, no. 5, pp. 1365-1372, 2019/05/24, 2019.

[62] T. B. Limbu, B. Chitara, M. Y. Garcia Cervantes, Y. Zhou, S. Huang, Y. Tang, and F. Yan, "Unravelling the Thickness Dependence and Mechanism of Surface-Enhanced Raman Scattering on Ti3C2TX MXene Nanosheets," *The Journal of Physical Chemistry C,* vol. 124, no. 32, pp. 17772-17782, 2020/08/13, 2020.

[63] K. Sakurai, and M. Mizusawa, "X-ray Diffraction Imaging of Anatase and Rutile," *Analytical Chemistry,* vol. 82, no. 9, pp. 3519-3522, 2010/05/01, 2010.

[64] D. B. Lioi, P. R. Stevenson, B. T. Seymour, G. Neher, R. D. Schaller, D. J. Gosztola, R. A. Vaia, J. P. Vernon, and W. J. Kennedy, "Simultaneous Ultrafast Transmission and Reflection of Nanometer-Thick Ti3C2Tx MXene Films in the Visible and Near-Infrared: Implications for Energy Storage, Electromagnetic Shielding, and Laser Systems," *ACS Applied Nano Materials,* vol. 3, no. 10, pp. 9604-9609, 2020/10/23, 2020.

[65] D. Jin, W. Liu, L. Jia, Y. Zhang, J. Hu, H. El Dirani, S. Kerdiles, C. Sciancalepore, P. Demongodin, C. Grillet, C. Monat, D. Huang, J. Wu, B. Jia, and D. J. Moss, "Thickness- and Wavelength-Dependent Nonlinear Optical Absorption in 2D Layered MXene Films," *Small Science,* vol. 4, no. 8, pp. 2400179, 2024.

[66] A. A. Shamsabadi, H. Fang, D. Zhang, A. Thakur, C. Y. Chen, A. Zhang, H. Wang, B. Anasori, M. Soroush, Y. Gogotsi, and Z. Fakhraai, "The Evolution of MXenes Conductivity and Optical Properties Upon Heating in Air," *Small Methods,* vol. 7, no. 10, pp. 2300568, 2023.

[67] L. Chen, M. Wakeel, T. U. Haq, C. Chen, and X. Ren, "Insight into UV-induced simultaneous photocatalytic degradation of Ti3C2Tx MXene and reduction of U(VI)," *Journal of Hazardous Materials,* vol. 430, pp. 128377, 2022/05/15/, 2022.

[68] Y. Zhang, J. Wu, Y. Yang, Y. Qu, L. Jia, T. Moein, B. Jia, and D. J. Moss, "Enhanced Kerr Nonlinearity and Nonlinear Figure of Merit in Silicon Nanowires Integrated with 2D Graphene Oxide Films," *ACS Applied Materials & Interfaces,* vol. 12, no. 29, pp. 33094-33103, 2020/07/22, 2020.

[69] Y. Yang, J. Wu, X. Xu, Y. Liang, S. T. Chu, B. E. Little, R. Morandotti, B. Jia, and D. J. Moss, "Invited Article: Enhanced four-wave mixing in waveguides integrated with graphene oxide," *APL Photonics,* vol. 3, no. 12, pp. 120803, 2018.

[70] G. E. Jellison, Jr., L. A. Boatner, J. D. Budai, B.-S. Jeong, and D. P. Norton, "Spectroscopic ellipsometry of thin film and bulk anatase (TiO2)," *Journal of Applied Physics,* vol. 93, no. 12, pp. 9537-9541, 2003.

[71] J. Wu, Y. Zhang, J. Hu, Y. Yang, D. Jin, W. Liu, D. Huang, B. Jia, and D. J. Moss, "2D graphene oxide films expand functionality of photonic chips," *Advanced Materials*, pp. 2403659, 2024.

[72] H. Arianfard, S. Juodkazis, D. J. Moss, and J. Wu, "Sagnac interference in integrated photonics," *Applied Physics Reviews,* vol. 10, no. 1, 2023.

[73] B. J. Eggleton, B. Luther-Davies, and K. Richardson, "Chalcogenide photonics," *Nature Photonics,* vol. 5, no. 3, pp. 141-148, 2011/03/01, 2011.

[74] S. Shekhar, W. Bogaerts, L. Chrostowski, J. E. Bowers, M. Hochberg, R. Soref, and B. J. Shastri, "Roadmapping the next generation of silicon photonics," *Nature Communications,* vol. 15, no. 1, pp. 751, 2024/01/25, 2024.

[75] A. H. Atabaki, S. Moazeni, F. Pavanello, H. Gevorgyan, J. Notaros, L. Alloatti, M. T. Wade, C. Sun, S. A. Kruger, H. Meng, K. Al Qubaisi, I. Wang, B. Zhang, A. Khilo, C. V. Baiocco, M. A. Popović, V. M. Stojanović, and R. J. Ram, "Integrating photonics with silicon nanoelectronics for the next generation of systems on a chip," *Nature,* vol. 556, no. 7701, pp. 349-354, 2018/04/01,

2018.
[76] C. Xiang, W. Jin, O. Terra, B. Dong, H. Wang, L. Wu, J. Guo, T. J. Morin, E. Hughes, J. Peters, Q.-X. Ji, A. Feshali, M. Paniccia, K. J. Vahala, and J. E. Bowers, "3D integration enables ultralow-noise isolator-free lasers in silicon photonics," *Nature,* vol. 620, no. 7972, pp. 78-85, 2023/08/01, 2023.
[77] J. T. Kim, and C.-G. Choi, "Graphene-based polymer waveguide polarizer," *Optics Express,* vol. 20, no. 4, pp. 3556-3562, 2012/02/13, 2012.
[78] C. Pei, L. Yang, G. Wang, Y. Wang, X. Jiang, Y. Hao, Y. Li, and J. Yang, "Broadband Graphene/Glass Hybrid Waveguide Polarizer," *IEEE Photonics Technology Letters,* vol. 27, no. 9, pp. 927-930, 2015.
[79] Z. Zhao, K. Tian, X. Li, L. Mao, S. Dai, P. Wang, and Y. Liu, "Scalable Mid-Infrared Polarization Control via Shape-Engineered Graphene-Integrated Waveguides," *Laser & Photonics Reviews,* vol. n/a, no. n/a, pp. e71176, 2026/04/07, 2026.
[80] N. Berahim, I. S. Amiri, T. Anwar, S. R. Azzuhri, M. N. S. Mohd Nasir, R. Zakaria, W. Y. Chong, C. K. Lai, S. H. Lee, H. Ahmad, M. A. Ismail, and P. Yupapin, "Polarizing effect of MoSe2-coated optical waveguides," *Results in Physics,* vol. 12, pp. 7-11, 2019/03/01/, 2019.
[81] Y. Tan, R. He, C. Cheng, D. Wang, Y. Chen, and F. Chen, "Polarization-dependent optical absorption of MoS2 for refractive index sensing," *Scientific Reports,* vol. 4, no. 1, pp. 7523, 2014/12/17, 2014.
[82] S. Sathiyan, H. Ahmad, W. Y. Chong, S. H. Lee, and S. Sivabalan, "Evolution of the Polarizing Effect of MoS2," *IEEE Photonics Journal,* vol. 7, no. 6, pp. 1-10, 2015.
[83] W. H. Lim, Y. K. Yap, W. Y. Chong, C. H. Pua, N. M. Huang, R. M. De La Rue, and H. Ahmad, "Graphene oxide-based waveguide polariser: From thin film to quasi-bulk," *Optics Express,* vol. 22, no. 9, pp. 11090-11098, 2014/05/05, 2014.
84. Moss, D. J., Morandotti, R., Gaeta, A. L. & Lipson, M. New CMOS compatible platforms based on silicon nitride and Hydex for nonlinear optics. Nat. Photonics Vol. 7, 597-607 (2013).
85. L. Razzari, et al., "CMOS-compatible integrated optical hyper-parametric oscillator," Nature Photonics, vol. 4, no. 1, pp. 41-45, 2010.
86. A. Pasquazi, et al., "Sub-picosecond phase-sensitive optical pulse characterization on a chip", Nature Photonics, vol. 5, no. 10, pp. 618-623 (2011).
87. M Ferrera et al., "On-Chip ultra-fast 1st and 2nd order CMOS compatible all-optical integration", Optics Express vol. 19 (23), 23153-23161 (2011).
88. Bao, C., et al., Direct soliton generation in microresonators, Opt. Lett, 42, 2519 (2017).
89. M.Ferrera et al., "CMOS compatible integrated all-optical RF spectrum analyzer", Optics Express, vol. 22, no. 18, 21488 - 21498 (2014).
90. M. Kues, et al., "Passively modelocked laser with an ultra-narrow spectral width", Nature Photonics, vol. 11, no. 3, 159 (2017).
91. M. Ferrera, et al., "Low-power continuous-wave nonlinear optics in doped silica glass integrated waveguide structures," Nature Photonics, vol. 2, no. 12, pp. 737-740, 2008.
92. M.Ferrera et al."On-Chip ultra-fast 1st and 2nd order CMOS compatible all-optical integration",

Opt. Express, vol. 19, (23)pp. 23153-23161 (2011).

93. D. Duchesne, M. Peccianti, M. R. E. Lamont, et al., “Supercontinuum generation in a high index doped silica glass spiral waveguide,” Optics Express, vol. 18, no, 2, pp. 923-930, 2010.

94. H Bao et al., “Turing patterns in a fiber laser with a nested microresonator: Robust and controllable microcomb generation”, Physical Review Research vol. 2 (2), 023395 (2020).

95. M. Ferrera, et al., “On-chip CMOS-compatible all-optical integrator”, Nature Communications, vol. 1, Article 29 (2010).

96. A. Pasquazi, et al., “All-optical wavelength conversion in an integrated ring resonator,” Optics Express, vol. 18, no. 4, 3858-3863 (2010).

97. A.Pasquazi, Y. Park, J. Azana, et al., “Efficient wavelength conversion and net parametric gain via Four Wave Mixing in a high index doped silica waveguide,” Optics Express, vol. 18, no. 8, pp. 7634-7641 (2010).

98. Peccianti, M. Ferrera, L. Razzari, et al., “Subpicosecond optical pulse compression via an integrated nonlinear chirper,” Optics Express, vol. 18, no. 8, pp. 7625-7633 (2010).

99. M. Ferrera et al., “All-optical 1st and 2nd order integration on a chip”, Optics Express vol. 19 (23), 23153-23161 (2011).

100. M. Ferrera et al., “Low Power CW Parametric Mixing in a Low Dispersion High Index Doped Silica Glass Micro-Ring Resonator with Q-factor > 1 Million”, Optics Express, vol.17, no. 16, 14098–14103 (2009).

101. M. Peccianti, et al., “Demonstration of an ultrafast nonlinear microcavity modelocked laser”, Nature Communications, vol. 3, 765, 2012.

102. A. Pasquazi, et al., “Self-locked optical parametric oscillation in a CMOS compatible microring resonator: a route to robust optical frequency comb generation on a chip,” Optics Express, vol. 21, no. 11, 13333-13341 (2013).

103. A. Pasquazi, et al., “Stable, dual mode, high repetition rate mode-locked laser based on a microring resonator,” Optics Express, vol. 20, no. 24, 27355-27362 (2012).

104. Pasquazi, A. et al. “Micro-combs: a novel generation of optical sources”, Physics Reports 729, 1-81 (2018).

105. H. Bao, et al., “Laser cavity-soliton microcombs”, Nature Photonics, vol. 13, no. 6, 384-389, (2019).

106. A. Cutrona et al., “High Conversion Efficiency in Laser Cavity-Soliton Microcombs”, Optics Express Vol. 30, Issue 22, 39816-39825 (2022).

107. M. Rowley et al., “Self-emergence of robust solitons in a micro-cavity”, Nature vol. 608 (7922) 303–309 (2022).

108. A. Cutrona et al., “Nonlocal bonding of a soliton and a blue-detuned state in a microcomb laser”, Nature Communications Physics 6 Article 259 (2023).

109. Aadhi A. et al., “Mode-locked laser with multiple timescales in a microresonator-based nested cavity”, APL Photonics 9 031302 (2024).

110. A. Cooper et al., “Parametric interaction of laser cavity-solitons with an external CW pump”, Optics Express 32 (12), 21783-21794 (2024).

111. A. Cutrona et al.,"Stability Properties of Laser Cavity-Solitons for Metrological Applications", Applied Physics Letters vol. 122 (12) 121104 (2023).
112. C. E. Murray et al., "Investigating the thermal robustness of soliton crystal microcombs", Optics Express 31(23), 37749-37762 (2023).
113. Y. Sun et al., "Enhancing laser temperature stability by passive self-injection locking to a micro-ring resonator", Optics Express 32 (13) 23841-23855 (2024).
114. Y. Sun et al., "Applications of optical micro-combs", Advances in Optics and Photonics 15 (1) 86-175 (2023).
115. X. Xu et al.,"Reconfigurable broadband microwave photonic intensity differentiator based on an integrated optical frequency comb source," APL Photonics, vol. 2, no. 9, 096104 (2017).
116. Xu, X., et al., Photonic microwave true time delays for phased array antennas using a 49 GHz FSR integrated micro-comb source, Photonics Research, vol. 6, B30-B36 (2018).
117. X. Xu et al., "Microcomb-based photonic RF signal processing", IEEE Photonics Technology Letters, vol. 31 no. 23 1854-1857 (2019).
118. A. Aadhi, L. Di Lauro, B. Fischer, P. Dmitriev, I. Alamgir, C. Mazoukh, N. Perron, E. Viktorov, A. Kovalev, A. Eshaghi, S. Vakili, M. Chemnitz, P. Roztocki, B.E. Little, S. T. Chu, D. J. Moss, and R. Morandotti, "Scalable Photonic Reservoir Computing for Parallel Machine Learning Tasks", Nature Communications 17 1-11 Article 1225 (2025). doi:10.1038/s41467-025-67983-z.
119. Haoran Zhang, Xiaotian Zhu, Xingyuan Xu, Shifan Chen, Yifu Xu, Jiajia Wang, Zhihui Liu, Shuai Wang, Yunping Bai, Chao Wang, Brent E. Little, Roberto Morandotti, David J. Moss, Sai T. Chu, Kun Xu, "Monolithic programmable microcombs", in press, Nature's Light: Science, and Applications (2026).
120. L. Peters, A. Cutrona, A. R. Cooper, L. Olivieri, F. Getman, V. Cecconi, N. Paul, D. Das, M. Rowley, S. T. Chu, B. E. Little, R. Morandotti, D. J. Moss, J. S. Totero Gongora, A. Pasquazi, M. Peccianti, "Millimetre-Wave Comb Generated by an Optical Microcomb", Nature Communications in press (2026).
121. Yang Li, Yang Sun, Jiayang Wu, Guanghui Ren, Thach G. Nguyen, Bill Corcoran, Xingyuan Xu, Sai T. Chu, Brent E. Little, Roberto Morandotti, Arnan Mitchell, and David J. Moss, "Reconfigurable Microwave Photonic Filters with Ultrasteep Roll-Off Based on Optical Microcombs", Laser and Photonics Reviews 20 e01910 (2026).
DOI: 10.1002/lpor.202501910.
122. Xingyuan Xu, et al., "Advanced adaptive photonic RF filters with 80 taps based on an integrated optical micro-comb source," Journal of Lightwave Technology, vol. 37, no. 4, 1288-1295 (2019).
123. X. Xu, et al., "Photonic RF and microwave integrator with soliton crystal microcombs", IEEE Transactions on Circuits and Systems II: Express Briefs, vol. 67, no. 12, 3582-3586 (2020).
124. X. Xu, et al., "High performance RF filters via bandwidth scaling with Kerr micro-combs," APL Photonics, vol. 4 (2) 026102 (2019).
125. M. Tan, et al., "Microwave and RF photonic fractional Hilbert transformer based on a 50 GHz Kerr micro-comb", Journal of Lightwave Technology, vol. 37, no. 24, 6097 – 6104 (2019).
126. M. Tan, et al., "RF and microwave fractional differentiator based on photonics", IEEE Transactions

on Circuits and Systems: Express Briefs, vol. 67, no.11, 2767-2771 (2020).
127. M. Tan, et al., "Photonic RF arbitrary waveform generator based on a soliton crystal micro-comb source", Journal of Lightwave Technology, vol. 38, no. 22, 6221-6226 (2020).
128. M. Tan et al., "RF and microwave high bandwidth signal processing based on Kerr Micro-combs", Advances in Physics X, VOL. 6, NO. 1, 1838946 (2021).
129. X. Xu, et al., "Advanced RF and microwave functions based on an integrated optical frequency comb source," Opt. Express, vol. 26 (3) 2569 (2018).
130. M. Tan et al., "Highly Versatile Broadband RF Photonic Fractional Hilbert Transformer Based on a Kerr Soliton Crystal Microcomb", Journal of Lightwave Technology vol. 39 (24) 7581-7587 (2021).
131. Wu, J. et al., "RF Photonics: An Optical Microcombs' Perspective", IEEE Journal of Selected Topics in Quantum Electronics Vol. 24, 6101020, 1-20 (2018).
132. T. G. Nguyen et al., "Integrated frequency comb source-based Hilbert transformer for wideband microwave photonic phase analysis," Opt. Express, vol. 23, no. 17, 22087-22097, (2015).
133. X. Xu, et al., "Broadband RF channelizer based on an integrated optical frequency Kerr comb source," Journal of Lightwave Technology, vol. 36, no. 19, pp. 4519-4526, (2018).
134. X. Xu, et al., "Continuously tunable orthogonally polarized RF optical single sideband generator based on micro-ring resonators," Journal of Optics, vol. 20, no. 11, 115701 (2018).
135. X. Xu, et al., "Orthogonally polarized RF optical single sideband generation and dual-channel equalization based on an integrated microring resonator," Journal of Lightwave Technology, vol. 36, no. 20, 4808-4818 (2018).
136. X. Xu, et al., "Photonic RF phase-encoded signal generation with a microcomb source", J. Lightwave Technology, vol. 38, no. 7, 1722-1727 (2020).
137. X. Xu, et al., "Broadband microwave frequency conversion based on an integrated optical micro-comb source", Journal of Lightwave Technology, vol. 38 no. 2, 332-338 (2020).
138. M. Tan, et al., "Photonic RF and microwave filters based on 49GHz and 200GHz Kerr microcombs", Optics Communications vol. 465,125563 (2020).
139. X. Xu, et al., "Broadband photonic RF channelizer with 90 channels based on a soliton crystal microcomb", Journal of Lightwave Technology, Vol. 38, no. 18, 5116 – 5121 (2020).
140. M. Tan et al, "Orthogonally polarized Photonic Radio Frequency single sideband generation with integrated micro-ring resonators", IOP Journal of Semiconductors, Vol. 42 (4), 041305 (2021).
141. M. Tan et al., "Photonic Radio Frequency Channelizers based on Kerr Optical Micro-combs", IOP Journal of Semiconductors Vol. 42 (4), 041302 (2021).
142. B. Corcoran, et al., "Ultra-dense optical data transmission over standard fiber with a single chip source", Nature Communications, vol. 11, Article:2568, (2020).
143. X. Xu et al, "Photonic perceptron based on a Kerr microcomb for scalable high speed optical neural networks", Laser and Photonics Reviews, vol. 14, no. 8, 2000070 (2020).
144. X. Xu, et al., "11 TOPs photonic convolutional accelerator for optical neural networks", Nature vol. 589, 44-51 (2021).
145. X. Xu et al., "Neuromorphic computing based on wavelength-division multiplexing", IEEE

Journal of Selected Topics in Quantum Electronics 29 (2) 7400112 (2023).
146. Y. Bai et al., "Photonic multiplexing techniques for neuromorphic computing", Nanophotonics vol. 12 (5): 795–817 (2023).
147. C. Prayoonyong et al., "Frequency comb distillation for optical superchannel transmission", Journal of Lightwave Technology vol. 39 (23) 7383-7392 (2021).
148. M. Tan et al., "Integral order photonic RF signal processors based on a soliton crystal micro-comb source", IOP Journal of Optics vol. 23 (11) 125701 (2021).
149. W. Han et al., "Dual-polarization RF Channelizer Based on Microcombs", Optics Express 32, No. 7, 11281-11295 (2024).
150. W. Han et al., Photonic RF Channelization Based on Microcombs", IEEE Journal of Selected Topics in Quantum Electronics 30 (5) 7600417 (2024).
151. X. Xu et al., "Microcomb-enabled parallel self- calibration optical convolution streaming processor", Light Science and Applications 15 149 (2026). doi.org/10.1038/s41377-025-02093-5.
152. Z. Liu et al., "Advances in Soliton Crystals Microcombs", Photonics Vol. 11, 1164 (2024).
153. B.Corcoran et al., "Optical microcombs for ultrahigh-bandwidth communications", Nature Photonics Volume 19 (5) 451 - 462 (2025).
154. S. Chen et al "Integrated photonic neural networks", npj Nanophotonics 2, 28 (2025).
155. Y. Li et al., "Feedback control in micro-comb-based microwave photonic transversal filter systems", IEEE Journal of Selected Topics in Quantum Electronics Vol. 30 (5) 2900117 (2024).
156. Y. Sun et al., "Optimizing the performance of microcomb based microwave photonic transversal signal processors", Journal of Lightwave Technology vol. 41 (23) 7223-7237 (2023).
157. M. Tan et al., "Photonic signal processor for real-time video image processing based on a Kerr microcomb", Nature Communications Engineering 2 94 (2023).
158. Y. Sun et al., "Quantifying the Accuracy of Microcomb-based Photonic RF Transversal Signal Processors", IEEE Journal of Selected Topics in Quantum Electronics vol. 29 no. 6, 1-17, 7500317 (2023).
159. C. Mazoukh et al., "Genetic algorithm-enhanced microcomb state generation", Nature Communications Physics Vol. 7, 81 (2024).
160. S. Chen et al., "High-bit-efficiency TOPS optical tensor convolutional accelerator using micro-combs", Laser & Photonics Reviews 19 2401975 (2025).
161. Y. Li et al., "Performance analysis of microwave photonic spectral filters based on optical microcombs", Advanced Physics Research 4 (9) 2400084 (2025).
162. L. di Lauro et al., "Optimization Methods for Integrated and Programmable Photonics in Next-Generation Classical and Quantum Smart Communication and Signal Processing", Advances in Optics and Photonics Vol. 17 (2) 526 - 622 (2025).
163. Y. Li et al., "Processing accuracy of microcomb-based microwave photonic signal processors for different input signal waveforms", Photonics 10, 10111283 (2023).
164. Y. Sun et al., "Comparison of microcomb-based RF photonic transversal signal processors implemented with discrete components versus integrated chips", Micromachines 14, 1794 (2023).
165. Chengzhuo Xia, Yifu Xu, Shifan Chen, Sirui Huang, Yunping Bai, Sai T. Chu, Brent E. Little,

Roberto Morandotti, David J. Moss, Xingyuan Xu, Kun Xu, " TOPS-speed Reconfigurable Photonic Transposed Convolution Accelerator for Generative Tasks", Laser & Photonics Reviews 20 e00771 (2026). https://doi.org/10.1002/lpor.202500771.

166. M. Tan et al., “The laser trick that could put an ultraprecise optical clock on a chip”, Nature 624, (7991) 256-257 (2023).
167. X. Yang et al., “Turnkey deterministic soliton crystal generation”, Laser and Photonics Reviews 19 (10) 2401687 (2025).
168. Y. Sun et al., “Self-locking of free-running DFB lasers to a single microring resonator for dense WDM”, Journal of Lightwave Technology 43, (4) 1995-2002 (2025).
169. W. Han et al., “TOPS-speed complex-valued convolutional accelerator for feature extraction and inference”, Nature Communications 16 292 (2025).
170. J. Hu et al., “Thermo-optic response and optical bistablility of integrated high index doped silica ring resonators”, Sensors 23 9767 (2023).
171. J. Hu et al., “Silicon photonic polarizers incorporating 2D MoS2 films”, Invited Paper, IEEE Journal of Selected Topics in Quantum Electronics 31 (2025). DOI:10.1109/JSTQE.2025.3610438
172. C. Khallouf et al., “Raman scattering and supercontinuum generation in high-index doped silica chip waveguides”, Nonlinear Optics and its Applications, edited by John M. Dudley, Anna C. Peacock, Birgit Stiller, Giovanna Tissoni, SPIE Vol. 13004, 130040I (2024).
173. M. Zerbib et al., “Observation of Brillouin scattering in a high-index doped silica chip waveguide”, Results in Physics 52 106830 (2023).
174. C. Khallouf et al., “Raman scattering and supercontinuum generation in high-index doped silica chip waveguides”, Nonlinear Optics and its Applications, edited by John M. Dudley, Anna C. Peacock, Birgit Stiller, Giovanna Tissoni, SPIE Vol. 13004, 130040I (2024).
175. C. Khallouf et al., “Supercontinuum generation in high-index doped silica photonic integrated circuits under diverse pumping settings”, Optics Express 33, 8431-8444 (2025).
176. C. Khallouf, L. Sader, A. Bougaud, G. Fanjoux, B. Little, S. T. Chu, D. J. Moss, R. Morandotti, G. P. Agrawal, J. M. Dudley, B. Wetzel, And T.Sylvestre, “Dual-pumping supercontinuum generation and temporal reflection in a nonlinear photonic integrated circuit”, Optics Express (2025).
177. A. Della Torre et al., “Mid-Infrared Supercontinuum Generation in a Varying Dispersion Waveguide for Multi-Species Gas Spectroscopy”, IEEE Journal of Selected Topics in Quantum Electronics 29 (1) 5100509 (2023).
178. Y. Zhang et al., “2D material integrated photonics: towards industrial manufacturing and commercialization”, Applied Physics Letters Photonics 10, 040903 (2025).
179. W. Jiang et al., “Enhanced thermo-optic performance for silicon microring resonators integrated with 2D graphene oxide films”, ACS Applied Electronic Materials 7 (12), 5650-5661 (2025).
180. Y. Yang et al., “Enhanced four-wave mixing in graphene oxide coated waveguides”, Applied Physics Letters Photonics vol. 3 120803 (2018).
181. Wu, J. et al., “Graphene oxide waveguide and micro-ring resonator polarizers”, Laser and Photonics Reviews Vol. 13, 1900056 (2019).
182. Y. Zhang et al., “Enhanced Kerr nonlinearity and nonlinear figure of merit in silicon nanowires

integrated with 2D graphene oxide films", ACS Applied Materials and Interfaces vol. 12 (29) 33094−33103 (2020).

183. Y. Qu et al., "Enhanced nonlinear four-wave mixing in silicon nitride waveguides integrated with 2D layered graphene oxide films", Advanced Optical Materials vol. 8 (21) 2001048 (2020).

184. Shahaz S. Hameed, Di Jin, Aihao Zhao, Jiayang Wu, Junkai Hu, Sebastien Cueff, Christian Grillet, Yuning Zhang, Irfan H. Abidi, Sumeet Walia, Christelle Monat, and David J. Moss, "Enhanced self-phase modulation in silicon nitride waveguides integrated with 2D MoS2 films", Advanced Materials Technologies 11 e02349 (2026). DOI: 10.1002/admt.202502349.

185. Rong Wang, Di Jin, Junkai Hu, Wenbo Liu, Yuning Zhang, Irfan H. Abidi, Sumeet Walia, Baohua Jia, Duan Huang, Jiayang Wu, and David J. Moss, "AI-guided design and optimization of 2D material based optical polarizers", Chip 5 (1) 100196 (2026). https://doi.org/10.1016/j.chip.2026.100196.

186. Rong Wang, Yijun Wang, Di Jin, Junkai Hu, Wenbo Liu, Yuning Zhang, Baohua Jia, Duan Huang, Jiayang Wu, and David J. Moss, "AI-guided optimization of integrated waveguide polarizers with 2D reduced graphene oxide", Journal of the Optical Society of America B (JOSA B) Vol. 43, No. 4, 793 – 803 (2026). https://doi.org/10.1364/JOSAB.583661

187. J. Wu et al., "Enhanced nonlinear four-wave mixing in microring resonators integrated with layered graphene oxide films", Small vol. 16 (16) 1906563 (2020).

188. J. Wu et al., "Graphene oxide waveguide polarizers and polarization selective micro-ring resonators", Paper 11282-29, SPIE Photonics West, San Francisco, CA, 4 - 7 February (2020).

189. Y. Zhang et al., "Design and optimization of four-wave mixing in microring resonators integrated with 2D graphene oxide films", Journal of Lightwave Technology Vol. 39 (20) 6553-6562 (2021).

190. Y. Qu et al., "Analysis of four-wave mixing in silicon nitride waveguides integrated with 2D layered graphene oxide films", Journal of Lightwave Technology Vol. 39 (9) 2902-2910 (2021).

191. J. Wu et al., "Graphene oxide: versatile films for flat optics to nonlinear photonic chips", Advanced Materials Vol. 33 (3) 2006415, 1-29 (2021).

192. Y. Qu et al., "Graphene oxide for enhanced optical nonlinear performance in CMOS compatible integrated devices", Paper No. 11688-30, PW21O-OE109-36, 2D Photonic Materials and Devices IV, SPIE Photonics West, San Francisco CA March 6-11 (2021). doi.org/10.1117/12.2583978

193. Y. Zhang et al., "Optimizing the Kerr nonlinear optical performance of silicon waveguides integrated with 2D graphene oxide films", Journal of Lightwave Technology Vol. 39 (14) 4671-4683 (2021).

194. Y. Qu et al., "Photo thermal tuning in GO-coated integrated waveguides", Micromachines Vol. 13 1194 (2022).

195. Zhang Y et al., "Graphene oxide-based waveguides for enhanced self-phase modulation", Annals of Mathematics and Physics Vol. 5 (2) 103-106 (2022). DOI:10.17352/amp.000048

196. Y. Zhang et al., "Enhanced spectral broadening of femtosecond optical pulses in silicon nanowires integrated with 2D graphene oxide films", Micromachines Vol. 13 756 (2022).

197. Y. Zhang et al., "Enhanced supercontinuum generated in SiN waveguides coated with GO films", Advanced Materials Technologies 8 (1) 2201796 (2023).

198. Y. Zhang et al.,"Graphene oxide for nonlinear integrated photonics", Laser and Photonics Reviews 17 2200512 (2023).
199. J. Wu et al., "Graphene oxide for electronics, photonics, and optoelectronics", Nature Reviews Chemistry 7 (3) 162–183 (2023).
200. Y. Zhang et al., "Enhanced self-phase modulation in silicon nitride waveguides integrated with 2D graphene oxide films", IEEE Journal of Selected Topics in Quantum Electronics Vol. 29 (1) 5100413 (2023).
201. Y. Qu et al., "Integrated optical parametric amplifiers in silicon nitride waveguides incorporated with 2D graphene oxide films", Light: Advanced Manufacturing 4 39 (2023).
202. J. Wu et al., "Novel functionality with 2D graphene oxide films integrated on silicon photonic chips", Advanced Materials Vol. 36 2403659 (2024).
203. D. Jin et al., "Silicon photonic waveguide and microring resonator polarizers incorporating 2D graphene oxide films", Applied Physics Letters, Vol. 125, 053101 (2024).
204. Y. Zhang et al., "Advanced optical polarizers based on 2D materials", npj Nanophotonics 1, 28 (2024).
205. J. Hu et al.,"2D graphene oxide: a versatile thermo-optic material", Advanced Functional Materials 34 2406799 (2024).
206. Y. Zhang et al., "Graphene oxide for enhanced nonlinear optics in integrated photonic chips", Paper 12888-16, Conference OE109, 2D Photonic Materials and Devices VII, Chair(s): Arka Majmdar; Carlos M. Torres Jr.; Hui Deng, SPIE Photonics West, San Francisco CA, January 27 – February 1 (2024). Proceedings Volume 12888, 2D Photonic Materials and Devices VII; 1288805 (2024). https://doi.org/10.1117/12.3005069
207. D. Jin et al., "Thickness and Wavelength Dependent Nonlinear Optical Absorption in 2D Layered MXene Films", Small Science 4 2400179 (2024).
208. J. Hu et al., "Integrated waveguide and microring polarizers incorporating 2D reduced graphene oxide", Opto-Electronic Science 4 240032 (2025).
209. L. Jia et al., "Third-order optical nonlinearities of 2D materials at telecommunications wavelengths", Micromachines, 14 307 (2023).
210. Linnan Jia, Jiayang Wu, Yuning Zhang, Yang Qu, Baohua Jia, Zhigang Chen, and David J. Moss, "Fabrication Technologies for the On-Chip Integration of 2D Materials", Small: Methods Vol. 6, 2101435 (2022).
211. L. Jia et al., "BiOBr nanoflakes with strong nonlinear optical properties towards hybrid integrated photonic devices", Applied Physics Letters Photonics vol. 4 090802 vol. (2019).
212. L. Jia et al "Large Third-Order Optical Kerr Nonlinearity in Nanometer-Thick PdSe2 2D Dichalcogenide Films: Implications for Nonlinear Photonic Devices", ACS Applied Nano Materials vol. 3 (7) 6876–6883 (2020).
213. Kues, M. et al. "Quantum optical microcombs", Nature Photonics vol. 13, (3) 170-179 (2019).
214. C.Reimer et al., "Integrated frequency comb source of heralded single photons," Optics Express, vol. 22, no. 6, 6535-6546, (2014).
215. C. Reimer, et al., "Cross-polarized photon-pair generation and bi-chromatically pumped optical

parametric oscillation on a chip", Nature Communications, vol. 6, 8236, (2015).
216. L. Caspani et al., "Multifrequency sources of quantum correlated photon pairs on-chip: a path toward integrated Quantum Frequency Combs," Nanophotonics, vol. 5, no. 2, pp. 351-362, (2016).
217. N. Montaut et al.,"Progress in integrated and fiber optics for time-bin based quantum information processing", Advanced Optical Technologies 14 1560084 (2025).
218. C. Reimer et al., "Generation of multiphoton entangled quantum states by means of integrated frequency combs," Science, vol. 351, no. 6278, pp. 1176-1180, 2016.
219. M. Kues, et al., "On-chip generation of high-dimensional entangled quantum states and their coherent control", Nature, vol. 546, no. 7660, pp. 622-626, 2017.
220. P. Roztocki et al., "Practical system for the generation of pulsed quantum frequency combs," Optics Express, vol. 25, no. 16, pp. 18940-18949, 2017.
221. Y. Zhang, et al., "Induced photon correlations through superposition of two four-wave mixing processes in integrated cavities", Laser and Photonics Reviews, vol. 14, no. 7, pp. 2000128, 2020.
222. C. Reimer, et al., "High-dimensional one-way quantum processing implemented on d-level cluster states", Nature Physics, vol. 15, no.2, pp. 148–153, 2019.
223. P.Roztocki et al., "Complex quantum state generation and coherent control based on integrated frequency combs", Journal of Lightwave Technology vol. 37 (2) 338-347 (2019).
224. S. Sciara et al., "Generation and Processing of Complex Photon States with Quantum Frequency Combs", IEEE Photonics Technology Letters vol. 31 (23) 1862-1865 (2019).
225. H. Yu et al., "Quantum key distribution implemented with d-level time-bin entangled photons", Nature Communications 16 171 (2025).
226. H. Yu et al., "Exploiting nonlocal correlations for dispersion-resilient quantum communications", Physical Review Letters 134 220801 (2025).
227. S. Sciara et al., "Scalable and effective multilevel entangled photon states: A promising tool to boost quantum technologies", Nanophotonics vol. 10 (18), 4447–4465 (2021).
228. L. Caspani et al., "Multifrequency sources of quantum correlated photon pairs on-chip: a path toward integrated Quantum Frequency Combs," Nanophotonics, vol. 5, no. 2, 351-362, 2016.
229. Stefania Sciara, Piotr Roztocki, Bennet Fisher, Christian Reimer, et al., "Scalable and effective multilevel entangled photon states: A promising tool to boost quantum technologies", Nanophotonics Vol. (11), 1-17 (2021).
230. H. Arianfard et al., "Sagnac interference in integrated photonics", Applied Physics Reviews 10 (1) 011309 (2023).
231. H. Arianfard et al., "Optical analogs of Rabi splitting in integrated waveguide-coupled resonators", Advanced Physics Research 2 2200123 (2023). DOI:10.1002/apxr.202200123.
232. H. Arianfard et al., "Spectral shaping based on optical waveguides with advanced Sagnac loop reflectors", Paper PW22O-OE201-20, SPIE-Opto, Integrated Optics: Devices, Materials, and Technologies XXVI, SPIE Photonics West, San Francisco CA January 22 - 27 (2022).
233. Di Jin et al., "Modelling of complex integrated photonic resonators using scattering matrix method", Photonics, Vol. 11, 1107 (2024).
234. H. Arianfard et al., "Spectral Shaping Based on Integrated Coupled Sagnac Loop Reflectors

Formed by a Self-Coupled Wire Waveguide", IEEE Photonics Technology Letters vol. 33 (13) 680-683 (2021).

235. H. Arianfard et al., "Three Waveguide Coupled Sagnac Loop Reflectors for Advanced Spectral Engineering", Journal of Lightwave Technology vol. 39 (11) 3478-3487 (2021).
236. H. Arianfard et al., "Advanced Multi-Functional Integrated Photonic Filters based on Coupled Sagnac Loop Reflectors", Journal of Lightwave Technology vol. 39 Issue: 5,1400-1408 (2021).
237. H. Arianfard et al., "Advanced multi-functional integrated photonic filters based on coupled Sagnac loop reflectors", Paper 11691-4, PW21O-OE203-44, Silicon Photonics XVI, SPIE Photonics West, San Francisco CA March 6-11 (2021).
238. J. Wu et al., "Advanced photonic filters via cascaded Sagnac loop reflector resonators in silicon-on-insulator integrated nanowires", Applied Physics Letters Photonics vol. 3 046102 (2018).
239. J. Wu et al., "Micro-ring resonator quality factor enhancement via an integrated Fabry-Perot cavity", Applied Physics Letters Photonics vol. 2 056103 (2017).
240. Hamed Arianfard, Jiayang Wu, Saulius Juodkazis, and David J. Moss, "Spectral shaping based on optical waveguides with advanced Sagnac loop reflectors", SPIE-Opto, Integrated Optics: Devices, Materials, and Technologies XXVI (2022). DOI
241. DJ Moss, HM van Driel, JE Sipe, "Dispersion in the anisotropy of optical third-harmonic generation in silicon", Optics letters Vol. 14 (1), 57-59 (1989).
242. D.J. Moss, J.E. Sipe, H.M. Van Driel, "Empirical tight-binding calculation of dispersion in the second-order nonlinear optical constant for zinc-blende crystals", Physical Review B36 (18), 9708 (1987).
243. D.J.Moss, S.Mclaughlin, G.Randall, M.Lamont, M. Ardekani, P.Colbourne, S.Kiran and C.A.Hulse, "Multichannel tunable dispersion compensation using all-pass multicavity etalons", Optical Fiber Communications Conference, Anaheim (2002) paper TuT2 page 132. Postconference Technical Digest (IEEE Cat. No.02CH37339). Opt Soc. America. Part vol.1, (2002), pp. 132-3. Washington, DC, USA.
244. L.M. Lunardi, D.J. Moss, S.Chandrasekhar, L.L.Buhl, A. Hulse, P.Colbourne, G.Randall, S.Mclaughlin, "Tunable dispersion compensators based on multi-cavity all-pass etalons for 40Gb/s systems", Journal of Lightwave Technology Vol. 20 (12) 2136 (2002). DOI: 10.1109/JLT.2002.806768
245. T Ido et al., "Strained InGaAs/InAlAs MQW electroabsorption modulators with large bandwidth and low driving voltage", IEEE photonics technology letters vol. 6 (10), 1207-1209 (1994).
246. D.J. Moss, HM Van Driel, JE Sipe, "Third harmonic generation as a structural diagnostic of ion-implanted amorphous and crystalline silicon", Applied physics letters vol. 48 (17), 1150-1152 (1986).
247. E. Ghahramani et al., "Second-harmonic generation in odd-period, strained, (Si(Ge/Si superlattices and at Si/Ge interfaces", Physical Review Letters vol. 64 (23), 2815 (1990).